\documentclass{SciPost}

\usepackage{graphicx}
\usepackage{hyperref}
\usepackage[bitstream-charter]{mathdesign}
\hypersetup{
    colorlinks,
    linkcolor={red!50!black},
    citecolor={blue!50!black},
    urlcolor={blue!80!black}
}

\DeclareSymbolFont{usualmathcal}{OMS}{cmsy}{m}{n}
\DeclareSymbolFontAlphabet{\mathcal}{usualmathcal}

\fancypagestyle{SPstyle}{
\fancyhf{}
\lhead{\colorbox{scipostblue}{\bf \color{white} ~SciPost Physics }}
\rhead{{\bf \color{scipostdeepblue} ~Submission }}

\fancyfoot[C]{\textbf{\thepage}}
}

\DeclareMathOperator{\tr}{tr}
\DeclareMathOperator{\Tr}{Tr}
\DeclareMathOperator{\const}{const}
\DeclareMathOperator{\sign}{sign}

\begin{document}

\pagestyle{SPstyle}

\begin{center}{\Large \textbf{
Mean-field theory of first-order quantum superconductor-insulator transition
}}\end{center}

\begin{center}
Igor Poboiko\textsuperscript{1,2$\star$} and
Mikhail V. Feigel'man\textsuperscript{3,4} 
\end{center}

\begin{center}
{\bf 1} 
Institute for Quantum Materials and Technologies, \\
Karlsruhe Institute of Technology (KIT),
76131 Karlsruhe, Germany
\\
{\bf 2} 
Institut f\"ur Theorie der Kondensierten Materie,\\
Karlsruhe Institute of Technology (KIT), 76131 Karlsruhe, Germany
\\
{\bf 3} Nanocenter CENN, Jamova 39, Ljubljana, SI-1000, Slovenia
\\
{\bf 4} LPMMC, University Grenoble-Alpes, Grenoble, France
\\

${}^\star$ \href{mailto:ihor.poboiko@kit.edu}{\small ihor.poboiko@kit.edu}
\end{center}

\begin{center}
    \today
\end{center}

\section*{Abstract}
{\bf
Recent experimental studies on strongly disordered indium oxide films have revealed an unusual first-order quantum phase transition between the superconducting and insulating states (SIT). This transition is characterized by a discontinuous jump from non-zero to zero values of superfluid stiffness at the critical point, contradicting the conventional ``scaling scenario'' typically associated with SIT. In this paper, we present a theoretical framework for understanding this first-order transition. Our approach is based on the concept of competition between two fundamentally distinct ground states that arise from electron pairs initially localized by strong disorder: the superconducting state and the Coulomb glass insulator. These ground states are distinguished by two crucially different order parameters, suggesting a natural expectation of a discontinuous transition between them at $T=0$. This transition occurs when the magnitudes of the superconducting gap $\Delta$ and the Coulomb gap $E_C$ become comparable. Additionally, we extend our analysis to low non-zero temperatures and provide a mean-field ``phase diagram'' in the plane of $(T/\Delta,E_C/\Delta)$. Our results reveal the existence of a natural upper bound for the kinetic inductance of strongly disordered superconductors.
}

\vspace{10pt}
\noindent\rule{\textwidth}{1pt}
\tableofcontents\thispagestyle{fancy}
\noindent\rule{\textwidth}{1pt}
\vspace{10pt}

\section{Introduction}

Studies of the transformation of a superconducting ground state into an insulating one (\textit{aka} Superconductor-Insulator Transition, or SIT for short)  
continue for more than four decades (see ~\cite{Review2020,Gantmakher-Dolgopolov,Goldman1998,Larkin1999,Fin1994,Sadovskii1997} 
for  various reviews of the subject) and still provide new discoveries. Very recent microwave resonance experiments~\cite{Exp-1st-order} on 
amorphous Indium Oxide films have made it possible to determine  superfluid stiffness $\Theta$ with a very high precision 
at low temperatures, and have demonstrated that superconductivity disappears (upon increase of disorder) in an abrupt manner.
Specifically, it was found that upon careful step-by-step increasing of disorder, a superconducting ground state 
with $\Theta > \Theta_{\text{min}} > 0$ suddenly transforms
into an insulating ground state with $\Theta=0$.
This result came as a surprise since all previously developed theories had considered it obvious that SIT is a kind of continuous quantum phase transition~\cite{cont-QPT}. In this paper, we will demonstrate that a self-consistent consideration of all major competing factors in the problem -- Anderson localization, superconducting pairing, and Coulomb interaction -- does indeed lead to the prediction of an abrupt first-order $T=0$ phase transition with a minimal
value of superfluid stiffness $\Theta_{\text{min}}$. Since $\Theta \propto 1/L_K$, where $L_K$ is the kinetic inductance of the film,
it means there is an upper bound  $L_K^{\text{max}}$ for the value of kinetic inductance of a superconductor.

 Amorphous InO$_x$ films serve as a prime example of a direct transition from a superconducting to an insulating state with increasing disorder~\cite{Shahar-Ovadyahu}. In the proximity of the Superconductor-Insulator Transition (SIT), these films exhibit a pseudogap in the single-particle density of states (DoS) that emerges well above the superconducting transition temperature ($T_c$) ~\cite{NaturePhys2011}. The origin of this pseudogap was explained in~\cite{FIKY,FIKC} by the binding energy ($\Delta_P$) between two electrons occupying the same localized state, akin to the parity gap observed in ultrasmall superconducting metal grains~\cite{Matveev-Larkin}. Crucially, this theory proposes that Cooper pairing occurs between Anderson-localized electrons, a possibility initially explored in~\cite{MaLee} and subsequently investigated numerically in~\cite{numerics1,numerics2} and analytically in~\cite{FIKY,FIKC}. The coexistence of a large single-particle gap ($\Delta_P$) with a much smaller collective superconducting gap ($\Delta$) was experimentally confirmed in ~\cite{Dubouchet} and corroborated by large-scale numerical simulations~\cite{numerics2011}. These findings collectively suggest that the low-temperature behavior of such superconductors (and the neighboring insulating state) can be effectively described using the ``pseudospin'' operators introduced by P.W. Anderson~\cite{Anderson-pseudospins}. This approach essentially neglects single-particle excitations as gapped entities, regardless of the superconducting state. In simpler terms, only a specific subspace of the entire Hilbert space, encompassing solely bound electron pairs, needs to be considered.

The theory of the disorder-driven SIT proposed in Refs.~\cite{IM-PRL,FIM2010}, based on a pseudospin Hamiltonian for Anderson-localized electrons, leads to a continuous quantum phase transition dominated by substantial statistical fluctuations. Near the transition point, the spatial fluctuations of the superconducting order parameter ($\Delta$) develop a "fat tail," making the concept of an average $\Delta$ meaningless. Consequently, the theory necessitates the use of a probability distribution, $P(\Delta)$, similar to the approach employed in a different context by Ref.~\cite{leDoussal}. However, a key limitation of the theory in Refs.~\cite{IM-PRL,FIM2010} is its neglect of the long-range Coulomb interaction between bound electron pairs.

The impact of Coulomb (or charging) energy on the SIT has been extensively studied in the context of artificial Josephson junction (JJ) arrays. A relevant review can be found in Ref.~\cite{Fazio}. In these systems, the SIT arises from the interplay between the Josephson coupling energy ($E_J$) between neighboring superconducting islands and the charging energy ($E_{\mathrm{charge}}$) associated with adding an extra Cooper pair to an island ($E_{\mathrm{charge}} = 2e^2/C$, where $C$ is the effective capacitance). While research has primarily focused on two-dimensional JJ arrays (three-dimensional granular arrays, although known~\cite{Gerber1997}, are more challenging to control), a definitive understanding of the SIT's nature in these systems remains elusive. Theoretically, the significant challenge lies in consistently accounting for the random stray charges inherent in such arrays. Experimentally, the behavior of JJ arrays near the SIT can be highly unconventional, with even hints of a "strange metal" state observed~\cite{strange,strange2}. However, amorphous InO$_x$ superconductors are fundamentally distinct from JJ arrays. Notably, they lack large, well-defined grains with a local order parameter (see Introduction to Ref.~\cite{FIKC} for a detailed discussion). Instead, the fundamental building block in amorphous InO$_x$ is the bound electron pair itself, aptly described by Anderson's pseudospin.

Diffusive transport theory offers a distinct approach to incorporating Coulomb interaction into the understanding of disorder effects on the critical temperature ($T_c$)~\cite{FinJap1,FinJap2,Fin1,Fin1994}. This theory, essentially an extension of BCS theory, accounts for the dependence of the effective electron-electron attraction on disorder, particularly when slow electron motion due to decreased diffusivity weakens the Coulomb repulsion. A.M. Finkelstein further developed~\cite{Fin1} and reviewed this approach~\cite{Fin1994}. Finkelstein's theory is well-suited when the suppression of $T_c$ by disorder goes hand-in-hand with a similar suppression of the spectral gap, a situation observed in many materials. However, amorphous InO$_x$, along with strongly disordered NbN and TiN films, exhibits a different behavior. Notably, Finkelstein's theory focuses on the short-range behavior of the Coulomb interaction, specifically within the coherence length (superconducting pair size)~\cite{FinJap1,FinJap2,Fin1,Fin1994}. Conversely, the competition between Josephson and charging energies discussed above involves a long-range Coulomb interaction acting between electron pairs.

It is noteworthy that the observed jump between states with non-zero and zero superfluid stiffness in Ref.~\cite{Exp-1st-order} bears a formal resemblance to the Berezinskii-Kosterlitz-Thouless (BKT) transition~\cite{BKT} known to occur in disordered superconducting thin films as a function of temperature. However, this similarity is superficial. The reported observation deals with a 3D phase transition (films are roughly 8-10 times thicker than the low-temperature coherence length, $\xi_0$) at near-zero temperature. This contrasts significantly with the BKT transition, which occurs in a 2D system driven by thermal fluctuations. BKT transitions are dominated by strong, long-wavelength fluctuations, which are absent in our case due to the near-zero temperature and higher dimensionality. Consequently, for the current problem, a mean-field approach like the Landau theory of phase transitions remains relevant, at least as a first approximation.

Section~\ref{sec:CoulombInt} delves into a detailed discussion of both short-range and long-range Coulomb interactions. We begin by demonstrating, based on the fractal nature of electron wavefunctions near the localization threshold~\cite{FIKC}, how phonon-induced electron-electron attraction can overcome Coulomb repulsion. This explains the existence of bound electron pairs despite Anderson localization. Subsequently, we address the magnitude of long-range Coulomb repulsion between electron pairs, incorporating recent experimental findings on insulating indium oxide films~\cite{Ebensperger,Scheffler2023} and the theoretical framework for the dielectric constant in Anderson insulators~\cite{FIC}. Our analysis reveals that the typical Efros-Shklovskii Coulomb gap~\cite{ES} ($E_C$) for near-critical InO$_x$ is estimated to be around 0.1 meV, comparable to the superconducting gap ($\Delta$). This observation forms the cornerstone of our theoretical approach presented in this work.

In Section~\ref{sec:Model}, we formulate a model of spin-1/2 pseudospins representing bound electron pairs. This model incorporates two key features: 1) an XY-type coupling (ferromagnetic-like) responsible for pair tunneling between localized orbitals, and 2) a long-range ZZ-type repulsion due to Coulomb interaction between the pairs. Random single-electron energies are modeled as random on-site "fields" acting along the Z pseudospin direction. The superconducting state is described by a ferromagnetic-like order parameter in the XY plane, while the insulating state corresponds to a random spin-glass-like ordering in the Z direction.

To understand the first-order Superconductor-Insulator Transition (SIT), a crucial observation is that this transition occurs between {\it two distinct ordered states} with fundamentally different order parameters. In simpler terms, due to the presence of long-range Coulomb repulsion, the insulating state possesses its own order parameter, a spin-glass order parameter of the Parisi type~\cite{MullerIoffe,MullerPankov}. This naturally leads to a discontinuous (first-order) phase transition: at the critical point, one order parameter vanishes simultaneously as the other one appears.

Since the first-order phase transition occurs at the point where the free energies of the two phases, the glassy insulating phase ($\mathcal{F}_{G}$) and the superconducting phase ($\mathcal{F}_{S}$), become equal, our primary goal is to calculate both free energies in the low-temperature limit and compare them. We will employ two different approximation methods.

First, in Section~\ref{sec:CG}, we will investigate the glassy insulating phase using a generalized Parisi approach. We will build upon the works by M\"uller, Pankov and Ioffe~\cite{MullerIoffe,MullerPankov} to calculate the free energy of this glassy state.
As expected in mean-field glass theories, the free energy of the replica symmetry breaking (RSB) state is predicted to be higher than that of the trivial (unstable) replica symmetric (RS) state.

Second, in Section~\ref{sec:SC}, we will focus on the superconducting phase when the Coulomb gap ($E_C$) is small compared to $\Delta_0$, the superconducting gap in the absence of Coulomb interaction. Here, we will calculate the free energy as a function of the superconducting order parameter ($\Delta$) and demonstrate that in the $T \rightarrow 0$ limit, it acquires an unusual non-analytic term proportional to $|\Delta| E_C$. This term effectively shifts the minimum of the energy functional for the order parameter $\Delta$ downwards compared to $\Delta_0$. For sufficiently large $E_C$, this non-trivial minimum disappears entirely. Even before that, when $E_C$ approaches $\Delta_0$, the ground state energy of the Coulomb glass becomes lower than that of the superconductor, inducing the SIT in this system.

In Section~\ref{sec:PhaseDiag}, we compare the free energies of the superconducting and glassy phases to determine the position of the phase transition line in the parameter space $(E_C/\Delta_0, T/\Delta_0)$. Finally, Section~\ref{sec:Discussion} presents a discussion of the results and conclusions. Some technical details are provided in Appendices~\ref{sec:appendix:MeanField}-\ref{sec:appendix:SC}.

\section{Coulomb interaction in near-critical Anderson insulators}
\label{sec:CoulombInt}

In this Section we provide several estimates for the strength of the Coulomb interaction in amorphous InO$_x$ films
near SIT. These estimates are not intended to be exact and systematic. Rather, the aim is to offer a new qualitative
picture and provide order-of-magnitude estimates which support it.

\subsection{Short length scales: competition between attraction and repulsion}
\label{sec:CoulombPhononCompetition}

This Section addresses a key question: why can InO$_x$ exhibit attractive electron-electron interactions that lead to superconductivity despite its high level of disorder, as measured by bulk resistivity? In contrast, superconductivity in many other bulk disordered materials, such as those described in Ref.~\cite{AndersonRamakrishnanMuttalib}, is suppressed to zero at significantly lower resistivity values.

Amorphous InO$_x$ stands out from many disordered superconductors due to its low density of conduction electrons ($n_e \sim 10^{21}~\text{cm}^{-3}$). These electrons interact through an insulating matrix with a relatively high dielectric constant ($\varepsilon$). However, it is important to note that the dielectric constant in Anderson insulators isn't a fixed value but depends on the length scale considered. 
A more accurate description of the dielectric response comes from the function $\varepsilon(q)$ in Fourier space. This function reaches a large macroscopic value ($\varepsilon$) at small wavevectors ($q \rightarrow 0$). According to Ref.~\cite{FIC}, the major contribution to this macroscopic $\varepsilon$ comes from large spatial scales ($\sim 5-10$ times the localization length, $\xi_{\text{loc}}$). This long-range $\varepsilon$ is crucial for understanding the interaction between localized electron pairs, which will be discussed later. 
On the other hand, to describe electron interaction within a single localized state (distances $\leq \xi_{\text{loc}}$), a much lower value of the short-range dielectric constant ($\varepsilon_1$) is needed. This $\varepsilon_1$ is expected to be roughly 50 or slightly higher (compared to the value of 30 reported for crystalline In$_2$O$_3$~\cite{zvi})

The effect of Coulomb interaction upon electron pairing in InO$_x$ was discussed in Sec.~1.2 of Ref.~\cite{FIKC} with the conclusion that $\varepsilon_1 \sim 30$ is sufficiently large to make overall Coulomb effect weak. Unfortunately, a subsequent error was identified in those calculations: the product $e^2k_F$ was significantly underestimated, being closer to 50,000 instead of 5,000 as originally stated~\cite{FIKC}. Following the same logic, this revision suggests a requirement of $\varepsilon_1 \geq 300$, which seems unrealistic.
However, there is another crucial factor that wasn't considered in the initial estimates: the fractal nature of electron wavefunctions at short scales ($ r \leq \xi_{\text{loc}}$). We will address this shortcoming below.
The key difference between the attractive and repulsive electron-electron interactions in electronic systems near the mobility edge lies in their range. Phonon-induced attraction is local, typically acting on the scale of a lattice constant. Conversely, Coulomb repulsion is long-range, with the interaction potential $U_C(r)$ decaying as $e^{2} / \varepsilon_1 r$. The fractal nature of electron wavefunctions ($\psi(\boldsymbol{r})$) enhances the matrix element of the local attractive interaction but has no effect on the long-range Coulomb repulsion.

In the following we will need an estimate for the Density of States (DoS) of strongly disordered InO$_x$, and we take it from
Ref.~\cite{Exp-1st-order}:
\begin{equation}
 \nu_0 \approx 1.2\cdot 10^{33}\, \text{erg}^{-1} \text{cm}^{-3}
\label{nu0}
\end{equation}  
Note that apparent difference with~\cite{Exp-1st-order} is due to the use of Gauss system of units in our paper; also we define here $\nu_0$ as the DoS per single projection of spin.

We now estimate average matrix element of local phonon-mediated
attraction of the form $U_{\text{e-ph}}(\boldsymbol{r} -\boldsymbol{r}^\prime) = - g_{\text{e-ph}} \delta(\boldsymbol{r} -\boldsymbol{r}^\prime)$, where
$g_{\text{e-ph}}$ is the electron-phonon coupling constant:
\begin{equation}
\overline{U_{\text{e-ph}}} = - \int d^3 \boldsymbol{r} d^3 \boldsymbol{r}^\prime g_{\text{e-ph}}\,\delta(\boldsymbol{r} -\boldsymbol{r}^\prime)\,
\overline{\psi^2(\boldsymbol{r})\psi^2(\boldsymbol{r}^\prime)}
\label{Ue1}
\end{equation}
To describe the correlation function of wavefunctions densities
$C(\boldsymbol{r},\boldsymbol{r}^\prime)  =  \overline{\psi^2(\boldsymbol{r})\psi^2(\boldsymbol{r}^\prime)} $
  we use scaling Ansatz~\cite{FIKC} valid for $d=3$:
\begin{eqnarray}
 C(\boldsymbol{r},\boldsymbol{r}^\prime)  =  
 \mathcal{I}_{3-d_{2}}^{-1}\, \xi_{\text{loc}}^{-6} \left(\frac{\xi_{\text{loc}}}{|\boldsymbol{r}-\boldsymbol{r}^\prime|}\right)^{3-d_2}
\exp\left( -\frac{|\boldsymbol{r}| + |\boldsymbol{r}^\prime|}{\xi_{\text{loc}}}\right),
\label{C1}
\end{eqnarray}
with the normalization constant given by integral:
\begin{equation}
{\cal I}_{\alpha}=\int\frac{d^{3}\boldsymbol{x}d^{3}\boldsymbol{y}e^{-|\boldsymbol{x}|-|\boldsymbol{y}|}}{|\boldsymbol{x}-\boldsymbol{y}|^{\alpha}}=\frac{4\pi^{2}}{3}(6-\alpha)(4-\alpha)(2-\alpha)\Gamma(2-\alpha)
\end{equation}
arising due to the condition $\int C(\boldsymbol{r},\boldsymbol{r}^\prime)d^3 \boldsymbol{r} d^3 \boldsymbol{r}^\prime =1$.
Here $d_2$ is the correlation fractal dimension of wavefunctions at the mobility edge; in 3D case $d_2 \approx 1.24$
according to results from numerical simulations~\cite{d22}. Distances $r$ and $r^\prime$ are measured from the center of localized wavefunction. For the use of $C(\boldsymbol{r}, \boldsymbol{r}^\prime)$  in Eq.\eqref{Ue1} one should set $|\boldsymbol{r} - \boldsymbol{r}^\prime| \to a$, which is short-distance cutoff of the order of lattice constant, to obtain
\begin{equation}
\overline{U_{\text{e-ph}}}\approx - \frac{\pi}{{\cal I}_{3-d_{2}}}\frac{g_{\text{e-ph}}}{a^3}\left(\frac{a}{\xi_{\text{loc}}}\right)^{d_{2}}
\label{Ue2}
\end{equation}
Note that small fractal exponent $d_2 $ makes the matrix element (\ref{Ue2}) much larger than it would be in the trivial case $d_2=3$, since the ratio $a/\xi_{\text{loc}} \ll 1$.

For the Coulomb matrix element, using the same wavefunctions and Eq.\eqref{C1}, we find:
\begin{eqnarray}
\overline{U_{C}}=\int d^{3}\boldsymbol{r}d^{3}\boldsymbol{r}^{\prime} \frac{e^{2}}{\varepsilon_{1}|\boldsymbol{r}-\boldsymbol{r}^{\prime}|}\overline{\psi^{2}(\boldsymbol{r})\psi^{2}(\boldsymbol{r}^{\prime})}=\frac{{\cal I}_{4-d_{2}}}{{\cal I}_{3-d_{2}}}\frac{e^{2}}{\varepsilon_{1}\xi_{\text{loc}}}
\label{UC}
\end{eqnarray}
Contrary to the case of local interaction, $\overline{U_C}$ in Eq.\eqref{UC} is nearly independent on $d_2$ and follows simplest
order-of-magnitude estimate.

Now we can estimate the ratio of between phonon-mediated attraction and  Coulomb repulsion:
\begin{equation}
\frac{\left|\overline{U_{\text{e-ph}}}\right|}{\overline{U_{C}}}\approx\lambda_0\,\frac{\pi}{{\cal I}_{4-d_{2}}}\left(\frac{a}{\xi_{\text{loc}}}\right)^{d_{2}-1}\frac{\varepsilon_{1}}{e^{2}\nu_{0}a^{2}}\approx2\lambda_0
\label{ratio}
\end{equation}
where  $\lambda_0 =  \nu_0 g_{\text{e-ph}}$ is the dimensionless Cooper attraction constant.
 For the last numerical estimate, we used Eq.\eqref{nu0} for $\nu_0$, then $a = 0.3~\text{nm}$,
 $\xi_{\text{loc}} = 5\,\text{nm}$, and $\varepsilon_1 = 50$.
The estimate (\ref{ratio}) shows that Coulomb repulsion can be overcome  
by phonon-induced attraction of moderate strength. The smallness of $d_2-1 = 0.24$ plays a crucial role in this result.
Indeed, with $d_2$ replaced by naive dimension $d=3$ one would get $\approx 0.01\lambda \ll 1$ in the R.H.S. of Eq.\eqref{ratio}, which would lead (incorrectly) to the conclusion that Coulomb repulsion dominates.

\subsection{Long length scales: Coulomb repulsion between bound pairs of electrons}
\label{sec:CoulombRepulsion}

The Coulomb repulsion between electron pairs tends to suppress fluctuations in the pair density, hindering the formation of long-range superconducting order. 
In our case, superconductivity emerges within an Anderson insulator with a relatively high density of states (DoS). Localized electrons with energies much larger than the superconducting gap ($|\epsilon| \gg \Delta$) do not contribute to superconductivity but act as a screening layer for the Coulomb interaction, effectively reducing the repulsion between bound pairs. A theoretical estimate for the resulting long-range dielectric constant ($\varepsilon$) in a 3D material without electron interaction was derived in Ref.~\cite{FIC}. For such a material with a full DoS of $2\nu_0$, the equation reads:
\begin{equation}
 \varepsilon \approx 40 \cdot 2 \nu_0 \cdot e^2\xi_{\text{loc}}^2
\label{diel}
\end{equation}
Equation~\eqref{diel} applies to relatively large localization lengths ($\xi_{\text{loc}}$) compared to the atomic scale (0.3 nm). However, its validity becomes questionable in the limit of infinitely large $\xi_{\text{loc}}$, where a different power-law dependence on $\xi_{\text{loc}}$ might emerge.
An important limitation of the calculation in Ref.~\cite{FIC} is that it neglects spatial inhomogeneity of the local electric field, particularly near the Anderson transition. This inhomogeneity could lead to a different critical exponent for the dependence of $\varepsilon$ on $\xi_{\text{loc}}$.
We expect Eq. \eqref{diel} to be reliable for a broad range of moderately large $\xi_{\text{loc}}$ in a non-interacting Anderson insulator. However, our focus is on an insulator with a pseudogap due to local electron-electron attraction. The modifications of dielectric response due to this feature were considered in Ref.~\cite{IF}. As a rough estimate, local pairing is expected to decrease $\varepsilon$ by a factor of about 2, which aligns with the data reported in Section 6.5.3 of the thesis~\cite{Ebensperger}.

Recent experimental data on dielectric constant in insulating amorphous InO$_x$ films with varying disorder levels support the above assertion~\cite{Ebensperger,Scheffler2023}. Notably, they observed that $\varepsilon$ measured using a microwave technique at ultra-low temperatures scales approximately as: $\varepsilon \propto T_0^{-\alpha}$, where $T_0$ is the activation temperature for transport conductivity ($\sigma(T) \propto e^{-T_0/T}$ in the same films). Theoretically we expect $\alpha = 2/d_2$  due to the relationship $T_0 \propto \xi_{\text{loc}}^{-d_2}$, as was shown in Ref.~\cite{FIKC},
and $\varepsilon \propto \xi_{\text{loc}}^2$ according to Eq.(\ref{diel}) above. The experimental value of $\alpha$ found in ~\cite{Ebensperger,Scheffler2023} ($\approx 1.54$) is in good agreement with this theoretical prediction. The maximum $\varepsilon$ measured in~\cite{Ebensperger,Scheffler2023} was around 500, with a corresponding $T_0 \approx 6$ K. It is reasonable to expect that the dielectric constant near the SIT could be even higher, reaching values of 1000-2000. This is because the lowest $T_0$ value measured in InO$_x$ films was around 2 K~\cite{T0}. On the other hand, using Eq. \eqref{diel} with typical values ($\nu_0 = 10^{33}~\text{erg}^{-1}\text{cm}^{-3}$ and $\xi_{\text{loc}} = 5~\text{nm}$), and accounting for the extra $1/2$ due to pairing pseudo-gap factor~\cite{IF}, we obtain $\varepsilon = 2000 $. This aligns with the extrapolation from the experimental data~\cite{Ebensperger,Scheffler2023}. For our numerical estimates of the Coulomb strength at long scales, we will use in the following an intermediate value of $\varepsilon = 1500$.

We will see below that the long-range repulsion
\begin{equation}
 U_{ij} = \frac{e^2}{\varepsilon |\boldsymbol{r}_i - \boldsymbol{r}_j|}
\label{U}
\end{equation}
between localized electron pairs (here $\boldsymbol{r}_{i,j} $  are centers of localized  wavefunctions $\psi_{i,j}$)
leads to a Coulomb gap like the one predicted by Efros and Shklovskii~\cite{ES}:
\begin{equation}
E_C = \sqrt{2\pi \nu_{0}} \frac{e^{3}}{\varepsilon^{3/2}}
\label{Ec3}
\end{equation}
We emphasise that the above Eq.\eqref{Ec3} serves just as the definition of the parameter $E_C$ for 3D Coulomb glass with $2e$ charges;
it differs by a numerical factor from similar quantity in Ref.~\cite{ES}.

While the main focus of this paper is on 3D disordered superconductors and thick films (thickness $d \gg \xi_0$), we expect the same phenomenon to occur in thin films with $d\ll \xi_0$ as well. The key difference lies in the magnitude of the Coulomb gap $E_C^{(\text{2D})}$ for thin films. To estimate $E_C^{(\text{2D})}$, we consider the very large dielectric constant $\varepsilon$, which leads to the electric field being concentrated within the film, extending up to distances on the order of $\sim \varepsilon d$ (typically in the micrometer range). Within this length scale, the Coulomb interaction energy becomes logarithmic, as described in Ref.~\cite{Keldysh-film}. In Fourier space, for electron pairs with a charge of $2e$, the 2D Coulomb interaction reads: $U_{C}^{(\text{2D})}(\boldsymbol{q}) = 4\pi e^2/\varepsilon d q^2$. The corresponding Coulomb gap value can be found in Ref.~\cite{Poboiko2020}:
\begin{equation}
    E_C^{(\text{2D})} = \frac{e^2}{\varepsilon d}
    \label{Ec2d}
\end{equation}
Equations \eqref{Ec3} and \eqref{Ec2d} represent the magnitudes of the Coulomb gap for bulk and 2D cases, respectively. These values will determine the positions of the first-order SIT in each scenario. 

It is worth noting that a unique situation exists in the superconductivity of the LaAlO$_3$/SrTiO$_3$ interface~\cite{La-STO}. Here, purely 2D electrons interact via bulk Coulomb forces, which are significantly suppressed due to the giant dielectric constant of SrTiO$_3$ (around $2\cdot 10^4$ at low temperatures). Consequently, the relevant $E_C$ in this case has been estimated to be negligibly small.

\section{Formulation of the model and general approach}
\label{sec:Model}

\subsection{Model Hamiltonian and general considerations}

We will use the Hamiltonian in the form (see Ref.~\cite{FIM2010} for comparison)
\begin{equation}
H = 2 \sum_i \xi_i S^z_i + \frac12 \sum_{i\neq j} \left[
-J_{ij} \left( S_i^+S_j^- + h.c.\right)
+ 4 U_{ij} S^z_i S^z_j 
\right]
\label{H0}
\end{equation}
where subscripts $i,j$ enumerate localized single-electron eigenstates, $\xi_i$ are their energies, factor 4 in front of $U_{ij}$ manifests $2e$ charge quantization, and
pseudospin operators are defined via electron creation/annihilation operators:
\begin{equation}
\label{spin-defs}
S_i^+ = a^\dagger_{i\uparrow}a^\dagger_{i\downarrow},  \qquad S_i^- = a_{i\downarrow}a_{i\uparrow}, \qquad
2S^z_i = a^\dagger_{i\uparrow}a_{i\uparrow} + a^\dagger_{i\downarrow}a_{i\downarrow} - 1,
\end{equation}
where $\uparrow,\downarrow$ denote possible directions of electron spin. 
Operators $S^\alpha$ from (\ref{spin-defs}) commute exactly as spin-$1/2$ operators~\cite{Anderson-pseudospins}.
Random local energies $\xi_i$ are uncorrelated and belong to a Gaussian wide band of width $W$, so that
 $P(\xi)=(\sqrt{2\pi}W)^{-1}\exp(-\xi^2 / 2 W^2)$.
Matrix elements $J_{ij}$ describe amplitudes of coherent tunneling of electron pairs between localized eigenstates:
\begin{equation}
J_{ij} = \frac{M}{ Z}\mathcal{A}_{ij}, 
\label{J}
\end{equation}
and are considered (like in Refs.~\cite{FIM2010,Anton1,Anton2}) to be equal for all ``connected'' pairs of states $i,j$,
 while $\mathcal{A}_{ij} $ is the connectivity matrix with  $Z \sim n_0 \xi_{\text{loc}}^3 \gg 1$ ``neighbours''  for each state $i$, with $n_0 \sim a^{-3}$ being spatial density of single-electron localized states.
Normalization of $J_{ij}$ is chosen such that the limit of $Z \to \infty$ is well-defined. The magnitude of $M$
can be related to effective microscopic electron attraction constant $g_{\text{eff}}$, which results
from the combination of attraction and repulsion couplings discussed in Sec. \ref{sec:CoulombPhononCompetition}, by the order-of-magnitude relation $M \sim g_\mathrm{eff} n_0$. Simultaneously, the energy eigenvalue density to be used below is
$P_0 = (\sqrt{2\pi} W)^{-1} = \nu_0/n_0 $.
Dimensionless coupling constant $\lambda = M P_0 = g_\mathrm{eff} \nu_0$ is always small in superconductors.
Below we will associate coordinates $\boldsymbol{r}_i$ with location of ``sites'' where our pseudospin operators $S^{\alpha}_i$
are located.  

A crucial simplification we will employ in this paper is that coordination number $Z$ is so large,
that local superconducting order parameter $\Delta_i$ is weakly space-dependent and can be replaced by a constant
$\Delta$.
According to Ref.~\cite{Anton1,Anton2}, it corresponds to the case of
\begin{equation}
\frac{\lambda W}{2\Delta Z} \ll 1.
\label{kappa}
\end{equation}
Here $\Delta \propto \exp(-1/\lambda)$ is very small, thus the condition (\ref{kappa}) is rather demanding.
Actually, experimental situation is such that upon increase of disorder the condition (\ref{kappa}) is
violated and spatial distribution of the superconducting order parameter becomes inhomogeneous~\cite{NaturePhys2011,Review2020}
even relatively far from the SIT. Nevertheless, we will use \eqref{kappa} in our further
considerations in this paper. We acknowledge it as a "mean-field approximation" that serves as a first step towards a more complete theory that incorporates superconductivity, localization, and Coulomb interaction.

The phase boundary between the superconducting and insulating states, assuming a discontinuous (first order) transition,
is determined by the condition where the free energies of both states are equal: $\mathcal{F}_{S} = \mathcal{F}_{G}$.
Our strategy will be to calculate these free energies in two distinct limits using a perturbative approach.

Our first step is to calculate the free energy of the glassy insulating state. We will assume that the dominant terms in the Hamiltonian (\ref{H0}) are the first and third terms. These terms likely represent the kinetic energy of the electrons and the Coulomb repulsion, respectively. This approach creates a formal similarity to the problem studied in Refs.~\cite{MullerIoffe,MullerPankov}, where purely classical Coulomb glasses were analyzed using the Parisi Replica-Symmetry-Breaking (RSB) approach. We will extend this approach in Section~\ref{sec:CG} to calculate the corresponding free energy for our system.

Second, we consider the opposite limit ($E_C \ll \Delta$) and calculate the free energy of the superconducting state in Section~\ref{sec:SC}, including corrections arising from Coulomb interaction. It is important to note that these corrections are absent in the classical theory of superconductivity for metals. This is because the Debye screening length in metals is typically on the atomic scale $a$, whereas the superconducting coherence length ($\xi_{0}$, or the size of a Cooper pair) is significantly larger than $a$. In clean superconductors, $\xi_0/a \sim E_F/\Delta \sim 10^3$. As a result, local electro-neutrality is maintained in metals with very high precision, and the formation of Cooper pairs doesn't affect the energy of Coulomb screening. However, the situation is entirely different for a superconducting state emerging within an Anderson insulator. Here, the relative magnitude of the energy correction is solely controlled by the ratio $E_C/\Delta$. As we will see, the non-trivial superconducting solution disappears when $E_C/\Delta \sim 1$.

\subsection{Mean-field approach}
To enable averaging over disorder and effectively describe the glassy phase of our model, we will employ the replica trick. This involves averaging the $n$-th power of the partition function, denoted as $\overline{Z^{n}}$, with the limit $n \to 0$ revealing the behavior of the average free energy. The detailed derivation of the mean field free energy functional is provided in Appendix \ref{sec:appendix:MeanField}. Below, we will briefly outline its key points and the approximations employed during its derivation.

One of the key observations that was noticed in Refs. \cite{MullerIoffe,MullerPankov} is that for large disorder strength $W$, the polarizability which governs the Debye screened Coulomb interaction $\hat{{\cal U}}$ via the RPA approximation 
\begin{equation}
\hat{{\cal U}}^{-1}=\hat{U}^{-1}+\hat{Q},
\end{equation}
becomes local in the real space and can be described by the matrix
\begin{equation}
\label{eq:Qmatrix:def}
Q_{i}^{ab}(\tau,\tau^{\prime})=4\left\langle \hat{S}_{i,a}^{z}(\tau)\hat{S}_{i,b}^{z}(\tau^{\prime})\right\rangle 
\end{equation}
Note that bare Coulomb interaction $\hat{U}$ is diagonal in the replica space and in Matsubara imaginary time 
$\tau \in (0,\beta=1/T)$.

Within the same approximation, the dynamics of local spin degrees of freedom is then governed by the single site effective action $\hat{A}_{\text{loc}}[\hat{G}_i,\hat{S}_i,\Delta_i]$ in which the effect of  long-range Coulomb interaction is incorporated into a ``Coulomb matrix'' $G_i^{ab}(\tau, \tau^\prime)$:
\begin{equation}
\label{eq:LocalSpinActionQuantum}
-\hat{A}_{\text{loc}}[\hat{G}, \hat{S}, \Delta]=2\int_{0}^{\beta}d\tau d\tau^{\prime}
\hat{S}_{a}^{z}(\tau)(W^{2}{\cal I}^{ab}+G^{ab}(\tau,\tau^{\prime}))\hat{S}_{b}^{z}(\tau^\prime)
+\int_{0}^{\beta}d\tau\sum_{a}(\Delta\hat{S}_{a}^{-}(\tau)+h.c),
\end{equation}
such that the correlation function \eqref{eq:Qmatrix:def} has to be calculated w.r.t. this action.

Finally, the ``Coulomb matrix''  $G_i^{ab}(\tau, \tau^\prime)$ can be shown to obey another equation, which closes self-consistency loop:
\begin{equation}
\label{eq:CoulombMatrix:def}
G_{i}^{ab}(\tau,\tau^{\prime})\equiv U_{ii}\delta^{ab}\delta(\tau-\tau^{\prime})-{\cal U}_{ii}^{ab}(\tau,\tau^{\prime}).
\end{equation}
In this relation, the first term simply cancels out unphysical self-action, while the second term describes the mean-field effect of the Coulomb interaction on a given spin  from the environment created by all other spins.

Assuming sites are uniformly distributed in space with with sufficiently large density $n_0$, one can switch to the coordinate basis and assume matrices to vary smoothly in space: $\hat{Q}_i = \hat{Q}(\boldsymbol{r}_i)$ and $\hat{G}_i = \hat{G}(\boldsymbol{r}_i)$. Furthermore, within the mean field approximation governed by the large coordination number $Z$, it suffices to consider spatially homogeneous configurations. The self-consistent procedure then can be conveniently formulated as the problem of finding the extremum of the mean-field free energy functional derived in Appendix \ref{sec:appendix:MeanField}:
\begin{equation}
\label{eq:MeanFieldFreeEnergy}
\mathcal{F}[\hat{G},\hat{Q},\Delta]=\mathcal{F}_C[\hat{G},\hat{Q}]+\mathcal{F}_\text{loc}[\hat{G},\Delta],
\end{equation}

The first contribution denoted as ``Coulomb'' is given by:
\begin{equation}
\label{eq:CoulombFreeEnergy}
    \mathcal{F}_C[\hat{G},\hat{Q}]=\frac{T}{2n}\Tr\left(n_{0}\hat{G}\hat{Q}-\Phi(\hat{Q})\right)
\end{equation}
where we have explicitly calculated the trace over the real space and left the trace over replica and imaginary time spaces:
\begin{equation}
\Phi(Q)=\int(d^3\boldsymbol{k})\left(n_{0}U_{\boldsymbol{k}}Q-\ln(1+n_{0}U_{\boldsymbol{k}}Q)\right)
=\frac{8}{3}\nu_{0}E_{C}\left(\frac{Q}{2 P_{0}}\right)^{3/2},
\end{equation}
with $E_C$ given by Eq. \eqref{Ec3}.

The second contribution, denoted as ``local'', is given by:
\begin{equation}
\label{eq:LocalFreeEnergy}
{\cal F}_{\text{loc}}[\hat{G}, \Delta]=
-\frac{T}{n} n_0\ln\Tr_{S}{\cal T}_{\tau}\exp\left(-\hat{A}_{\text{loc}}[\hat{G},\hat{S},\Delta]\right)
+\frac{\nu_{0}|\Delta|^{2}}{\lambda},
\end{equation}
with trace taken over quantum spin degrees of freedom, and symbol ${\cal T}_\tau$ denoting imaginary time ordering. The variation of the total free energy w.r.t. $\hat{G}$ then reproduces Eq. \eqref{eq:Qmatrix:def} and \eqref{eq:CoulombMatrix:def} correspondingly. Below we will analyze the behavior of this functional in various limits.

\section{Glassy insulating phase}
\label{sec:CG}
We start our treatment of the mean field model \eqref{eq:MeanFieldFreeEnergy} by first focusing on the Coulomb glass phase, i.e. neglecting all superconducting correlations: $\Delta = 0$. The analysis is then greatly simplified as the absence of the transverse field for the local spin action \eqref{eq:LocalSpinActionQuantum} makes it classical, allowing us to replace spin variables by binary $S = \pm 1/2$ and restrict ourselves to the zeros Matsubara harmonic only $Q_{ab}(\tau,\tau^{\prime}) = Q_{ab} = \const$, and same for $G^{ab}(\tau, \tau^\prime)$. The problem then becomes equivalent to one studied in Refs. \cite{MullerPankov,MullerIoffe}. The trace over imaginary time in Eq. \eqref{eq:CoulombFreeEnergy} yields additional factors of $\beta$, leading us to the following expression:
\begin{equation}
{\cal F}_{C}[\hat{G},\hat{Q}]=\frac{T}{2n}{\rm Tr}\left(n_{0}\beta^{2}\hat{G}\hat{Q}-\Phi(\beta\hat{Q})\right),
\end{equation}
with the remaining trace taken only over the replica space, whereas the ``action'' governing the local contribution to free energy via Eq. \eqref{eq:LocalFreeEnergy} reduces to:
\begin{equation}
-A_{\text{loc}}[G,S,\Delta=0]=2 \beta^2S_{a}(W^{2}{\cal I}^{ab}+G^{ab})S_{b}
\label{A1loc}
\end{equation}

At this point it becomes convenient to introduce a dimensionless inverse temperature, polarizability and dimensionless Coulomb matrix:
\begin{equation}
b \equiv \beta E_C,\quad \hat{q} \equiv \beta \hat{Q} / 2 P_0,\quad \hat{g} \equiv \hat{G} / 2 E_C^2,
\end{equation}
and the self-consistency equation then becomes:
\begin{equation}
\label{eq:Coupling}
\beta\hat{G}=\Phi^{\prime}(\beta \hat{Q}) / n_0\Rightarrow \hat{g} = \hat{q}^{1/2} / b,
\end{equation}
with the help of which the Coulomb contribution simplifies to:
\begin{equation}
{\cal F}_{C}=\frac{T}{2n}\Tr\left(\beta\hat{Q}\Phi^{\prime}(\beta\hat{Q})-\Phi(\beta\hat{Q})\right)
=\nu_{0}E_{C}T\,\frac{1}{n}\Tr\left(\frac{2}{3}\hat{q}^{3/2}\right).
\end{equation}

We utilize standard Parisi ultrametric ansatz as described in Appendix \ref{sec:appendix:RSB}, and parametrize ultrametric matrices $\hat{G}$ and $\hat{Q}$ via functions $G(x \in [0,1])$ and $Q(x \in [0,1])$, while their corresponding replica Fourier transforms are marked with tilde. We arrive at:
\begin{equation}
\label{eq:CoulombParisiResult}
\mathcal{F}_C = \frac{1}{2}n_{0}E_{C}-\frac{1}{3}\nu_{0}E_{C}T
+\nu_{0}E_{C}T\int_{0}^{1}dx\left(1-\frac{1}{x}\right)\widetilde{q}^{1/2}(x)\widetilde{q}^{\prime}(x)
\end{equation}

Now we switch our attention to the ``local'' contribution to the free energy. First of all, it contains a large band contribution:
\begin{equation}
\label{eq:BandFreeEnergy}
{\cal F}_b=-n_{0}T\int_{-\infty}^{+\infty} d\xi\,P_{0}(\xi)\ln2\cosh\beta\xi
\approx-\sqrt{\frac{2}{\pi}}n_{0}W-\frac{\pi^{2}}{12}\nu_{0}T^{2},
\end{equation}
after subtracting which the remaining contribution becomes dominated by the vicinity of the Fermi surface allowing one to replace the distribution function by a constant $P_0(\xi) \approx P_0$. The remaining part can be calculated with the help of Parisi scheme, the detailed analysis of which is presented in Appendix \ref{sec:appendix:Parisi}:
\begin{equation}
\label{eq:LocalParisiResult}
{\cal F}_{\text{loc}}-{\cal F}_{b}=\nu_{0}E_{C}^{2}\int_{-\infty}^{+\infty}dy\,y\left(m(0,y)-m(1,y)\right)
-\frac{3}{2}n_{0}E_{C}+\nu_{0}E_{C}T
\end{equation}

The magnetization $m(x,y)$ (with dimensionless field $y = \xi / E_C$), together with the dimensionless distribution function of local fields $p(x,y)$ and glass order parameter $\widetilde{q}(x)$ is calculated with the help of full set of Parisi equations:
\begin{equation}
\label{eq:ParisiM}
-\partial_{x}m=g^{\prime}(x)\left(\partial_{y}^{2}m+bx\partial_{y}(m^{2})\right), 
\end{equation}
where $m(x=1,y)=\tanh by$,
\begin{equation}
\label{eq:ParisiP}
\partial_{x}p=g^\prime(x)\left(\partial_{y}^{2}p-2bx\partial_{y}(mp)\right),
\end{equation}
where $p(x=0,y)=1$, and
\begin{equation}
\label{eq:ParisiQ}
\widetilde{q}(x)=\frac{1}{2}\int_{-\infty}^{+\infty} dy\,p(x,y)\partial_{y}m(x,y).
\end{equation}
The self-consistency loop is then closed with the help of Eq. \eqref{eq:Coupling} and replica Fourier transform:
\begin{equation}
\label{eq:ParisiG}
g^{\prime}(x)\equiv-\frac{\widetilde{g}^{\prime}(x)}{x}=-\frac{\widetilde{q}^\prime(x)}{2 b x \widetilde{q}^{1/2}(x)}.
\end{equation}

The Eqs. \eqref{eq:ParisiM}-\eqref{eq:ParisiG} constitute a dimensionless closed self-consistent scheme for describing the glass state of the classical Coulomb Glass and suitable for further numerical analysis; it depends on a single dimensionless parameter $b = \beta E_C$.

These equations always have a replica-symmetric solution with $\widetilde{q}(x) = 1$ and all quantities independent on $x$; this solution, however, becomes unstable at the Almeida-Thouless critical line where the marginal stability condition is fulfilled:
\begin{equation}
\label{eq:MarginalStability}
\widetilde{q}^{1/2}(x)=\frac{1}{2}\int_{-\infty}^{+\infty} dy\, p(x,y)\left(\partial_y m(x,y)\right)^{2},
\end{equation}
which yields the following glass transition temperature:
\begin{equation}
\label{eq:G:Temp}
T_G = \frac{2}{3} E_C.
\end{equation}

The constant contributions in Eq. \eqref{eq:CoulombParisiResult} and \eqref{eq:LocalParisiResult} are then combined into the free energy of the normal state (unstable at $T < T_G$):
\begin{equation}
\label{eq:N:FreeEnergy}
\mathcal{F}_N =-n_{0}\left(\sqrt{\frac{2}{\pi}}W+E_{C}\right)+\frac{2}{3}\nu_{0}E_{C}T-\frac{\pi^{2}}{12}\nu_{0}T^{2},
\end{equation}
while the difference of the free energy of glass state and the normal state is given by:
\begin{equation}
\label{eq:G:FreeEnergy}
({\cal F}_{G}-{\cal F}_{N})/\nu_{0}E_{C}^{2}=2\int_{0}^{1}dx\left(1-x\right)\widetilde{q}(x)g^{\prime}(x)
-\int_{-\infty}^{\infty}dy\,y\left(m(1,y)-m(0,y)\right)
\end{equation}

Last but not least, the entropy density $S_G = -\partial\mathcal{F}_G / \partial{T}$ of the glass state can also be calculated as:
\begin{equation}
\label{eq:G:Ent}
S_{G}/\nu_{0}E_{C}=\int_{-\infty}^{+\infty} dy\,p(1,y)\left[\ln2\cosh by-by\tanh by\right]
-\frac{2}{3}\widetilde{q}^{3/2}(1)
\end{equation}

\begin{figure}
    \includegraphics[width=\textwidth]{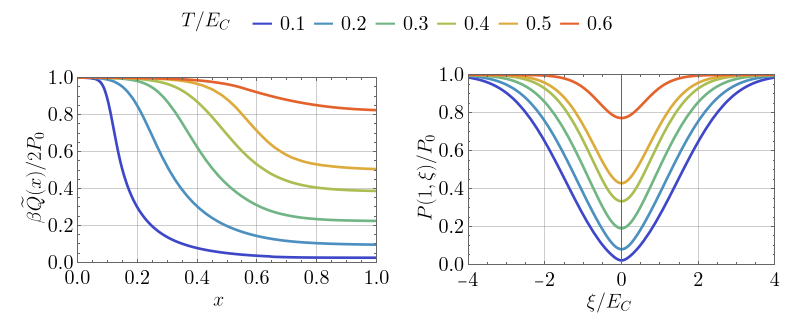}
    \centering
    \caption{Left: RSB order parameter function $\widetilde{Q}(x)$, right: distribution function of frozen potential $P(x=1, \xi)$, plotted for different values of $T / E_C$, obtained from numerical solution of Parisi equations.}
    \label{fig:G:Solution}
\end{figure}

\begin{figure}
\includegraphics[width=\textwidth]{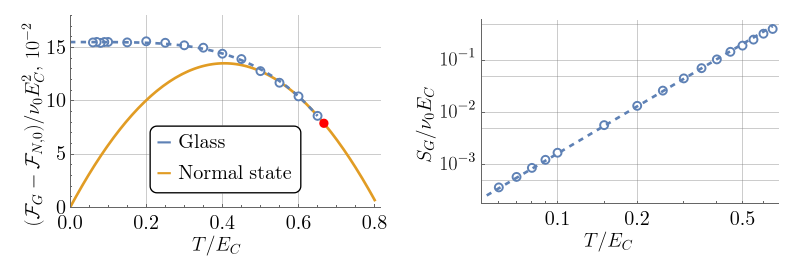}
\centering
\caption{Left: free energy of a glass state (blue), compared to the free energy of the normal state (orange). Red point marks the glass transition $T_G / E_C = 2/3$. Right: low-temperature behavior of entropy of glass state, dashed line: asymptotic behavior as given by Eq. \eqref{eq:G:Ent:LowT}.}
\label{fig:G:FreeEnergy}
\end{figure}

We have implemented the numerical iterative solution of the self-consistent scheme described by Eqs. 
\eqref{eq:ParisiM}-\eqref{eq:ParisiG}
for temperatures $T / E_C \in [0.06, 2/3]$ and have calculated the free energy associated with such temperatures. Fig. \ref{fig:G:Solution} demonstrates the characteristic behavior of RSB order parameter function $\widetilde{Q}(x)$ and distribution function of frozen fields $P(x=1, \xi)$ for several selected values of temperature; the latter corresponds physically to the tunneling density of states in Coulomb glass insulator.
In Fig. \ref{fig:G:FreeEnergy} we plot the temperature dependence of the obtained free energy as compared to the zero-temperature limit of free energy of the normal state $\mathcal{F}_{N,0} \equiv \mathcal{F}_N(T = 0)$, and the entropy of the glass state.

At low temperature $T / E_C \to 0$, solution to Parisi equations acquire a universal scaling form (see Ref. \cite{MullerPankov} and Appendix \ref{sec:appendix:Pankov}), 
leading to the appearance of the soft Coulomb gap in the density of states:

\begin{equation}
\label{eq:CoulombGap}
p(1,y)\simeq0.327\,y^2+2.298\,(T/E_{C})^{2},\quad T,\xi\ll E_{C}
\end{equation}
Numerical coefficients provided above are consistent with those obtained in Ref. \cite{MullerPankov}.
The entropy of the glass state, as given by Eq. \eqref{eq:G:Ent}, vanishes at $T \to 0$ as 
\begin{equation}
    \label{eq:G:Ent:LowT}
    S_G / \nu_0 E_C \approx 1.657\,(T / E_C)^3,\quad T \ll E_C
\end{equation}
We conclude in passing that for the similar reasons the entropy of 2D Coulomb glass with $1/r$ interaction scales as $S_G^{(\text{2D})}/\nu_0 E_C \propto (T/E_C)^2$. Note also that the case of 2D Coulomb glass with logarithmic interaction~\cite{Poboiko2020} is quite different, and careful analysis of its low-$T$ thermodynamics will be provided in a separate study.

\section{Superconducting phase: mean field solution corrected for Coulomb effects}
\label{sec:SC}

In the Section we consider situation of Coulomb effects weak compared to superconducting pairing, thus it is
sufficient to work within replica-symmetric manifold for $\hat{Q}$-matrix. For the same reason, it is possible
to expand the free energy (\ref{eq:MeanFieldFreeEnergy},\ref{eq:LocalFreeEnergy}) up to linear order in $\hat{G}$.
Variation of local free energy (\ref{eq:LocalFreeEnergy}) over $\hat{G}$ is equal to $\hat{Q}/2$, thus this term cancels out with
the corresponding term in the first line in Eq.\eqref{eq:MeanFieldFreeEnergy}, so that $\hat{G}$ does not enter at all within the
necessary accuracy. The details of the calculation are outlined in the Appendix \ref{sec:appendix:SC}.

The ``local'' contribution in the replica-symmetric case is given by:
\begin{equation}
\label{eq:SC:Floc}
{\cal F}_{\text{loc}}=n_0\int d\xi P_{0}(\xi)\left(-T\ln2\cosh\beta\sqrt{\xi^{2}+|\Delta|^{2}}\right).
\end{equation}
It is convenient to subtract the zero-temperature ``band'' contribution \eqref{eq:BandFreeEnergy}; the remaining integral is dominated by the vicinity of the Fermi energy allowing one to replace the distribution function by a constant. It also contains logarithmic divergence associated with the Cooper instability, and has to be cut at Debye frequency $\omega_D$. We then obtain:
\begin{equation}
\label{eq:SC:LocalContribution}
{\cal F}_{\text{loc}}(\Delta)-{\cal F}_{b,0}=-\nu_{0}|\Delta|^{2}\left(\frac{1}{2}\ln\frac{e \Delta_0^2}{|\Delta|^2}+\eta(\beta|\Delta|)\right),
\end{equation}
with the dimensionless function 
\begin{equation}
\label{eq:eta}
\eta(z)=\frac{1}{z}\int_{-\infty}^{\infty}dy\ln\left(1+e^{-2z\sqrt{y^{2}+1}}\right)
\end{equation}
and $\Delta_0 = 2 \omega_D \exp(-1 / \lambda)$. This is the only contribution in the absence of Coulomb energy, and its minimization over $\Delta$ leads to (nearly) standard result for $\Delta(T)$ dependence, up to
the replacement $2T \to T$ due to the reduced Hilbert space of our model (no single-particle excitations).

We now switch to the Coulomb corrections. In the replica-symmetric case, the $Q$-matrix contains just two components. The first one is the analog of Edwards-Anderson order parameter, it is purely static and characterizes the quenched fluctuations of single electron orbital occupation, i.e. magnetization in the spin language:
\begin{equation}
m(\xi) \equiv 2\langle S^z \rangle = \frac{\xi}{\sqrt{\xi^2+|\Delta|^2}}\tanh\beta\sqrt{\xi^2+|\Delta|^2},
\label{m1}
\end{equation}
and is given by:
\begin{equation}
\label{eq:SC:Q0}
Q_{0}=\int_{-\infty}^{+\infty} d\xi\,P_{0}(\xi)m^{2}(\xi).
\end{equation}

The second component of $Q$-matrix is associated with the dynamic screening and can be expressed via the Fourier transform of the dynamical spin susceptibility:
\begin{equation}
\chi(\xi,\tau)=4\left\langle \left\langle \hat{S}^{z}(\tau)\hat{S}^{z}(0)\right\rangle \right\rangle,
\end{equation}
as follows:
\begin{equation}
\label{eq:SC:Q1}
\widetilde{Q}_{1}(\omega)=\int_{-\infty}^{+\infty} d\xi\,P_{0}(\xi)\chi(\xi,\omega).
\end{equation}

The susceptibility at nonzero Matsubara frequency $\omega \neq 0$ is given by:
\begin{equation}
\chi(\xi,\omega\neq0)=\frac{|\Delta|^{2}\tanh\beta\sqrt{\xi^{2}+|\Delta|^{2}}}{\sqrt{\xi^{2}+|\Delta|^{2}}\left(\xi^{2}+|\Delta|^{2}+(\omega/2)^{2}\right)},
\end{equation}
and $\chi(\xi,\omega=0)=m^{\prime}(\xi)$. 

The Coulomb energy \eqref{eq:CoulombFreeEnergy} in the replica-symmetric limit is then expressed via components of $Q$-matrix as follows:
\begin{equation}
\label{eq:SC:FC}
{\cal F}_{C}=-\frac{1}{2}Q_{0}\Phi^{\prime}(\widetilde{Q}_{1}(0))-\frac{T}{2}\sum_{\omega}\Phi(\widetilde{Q}_{1}(\omega)),
\end{equation}
with the summation performed over bosonic Matsubara frequencies $\omega = 2 \pi n T$.

The resulting free energy of the superconducting phase can be written as:
\begin{equation}
\label{eq:SC:FreeEnergy}
{\cal F}_{S}-{\cal F}_{N,0}=\nu_{0}E_{C}|\Delta|\left(q_{0}(\beta|\Delta|)-\frac{4}{3}c(\beta|\Delta|)\right)
-\nu_{0}|\Delta|^{2}\left(\frac{1}{2}\ln\frac{e \Delta_{0}^2}{|\Delta|^2} + \eta(\beta|\Delta|)\right)
\end{equation}
where we have subtracted the normal state free energy at zero temperature $\mathcal{F}_{N,0} = \mathcal{F}_N(T=0)$, see Eq. \eqref{eq:N:FreeEnergy}, and dimensionless functions $q_0(z)$ and $c(z)$ are defined in Appendix \ref{sec:appendix:SC}. In the low temperature limit it takes a simple yet non-analytic in $\Delta$ form:
\begin{equation}
\label{eq:SC:FS}
{\cal F}_{S}(\Delta, T \to 0) - {\cal F}_{N, 0}\approx 1.947 \,\nu_{0}E_{C}|\Delta|-\frac{1}{2}\nu_{0}|\Delta|^{2}\ln\frac{e\Delta_{0}^{2}}{|\Delta|^{2}}.
\end{equation}
First term in \eqref{eq:SC:FS} comes due to decrease of ZZ susceptibility $\chi(\xi)$ 
in presence of the ``spin'' ordering in XY plane, described by $\Delta$.

\begin{figure}
    \centering
    \includegraphics[width=0.6\textwidth]{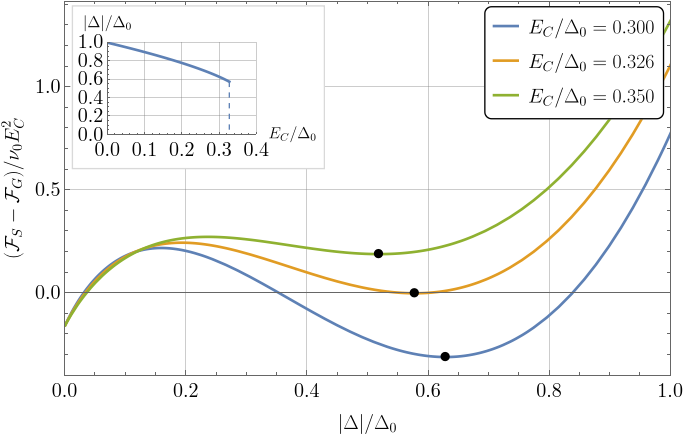}
    \caption{$\Delta$-dependence of free energy difference between superconducting free energy at zero temperature, given by Eq. \eqref{eq:SC:FS}, and glass free energy as given by Eq. \eqref{eq:G:FreeEnergy}, for different values of $E_C / \Delta_0$. The positions of minimum w.r.t. $\Delta$ are marked by black dots. Inset: $|\Delta|$ as a function of $E_C$ at zero temperature. First order transition occurs at $E_C \approx 0.326 \Delta_0$, corresponding to the orange curve, with $|\Delta|$ jumping from 0 to $|\Delta| = 0.577 \Delta_0$.}
    \label{fig:FSC}
\end{figure}

Equation (\ref{eq:SC:FS}) is the central result of this work:
 is describes competition between Coulomb gap and superconductivity.
This competition results in the decrease of magnitude of $\Delta$ which minimizes the energy \eqref{eq:SC:FS} with respect to the bare value $\Delta_0$. Figure \ref{fig:FSC} displays several curves for $\mathcal{F}_{S}(\Delta, T\to 0) - \mathcal{F}_{G} (T\to 0) $ at different values of the ratio $E_C/\Delta_0 < 1$.

At the critical value of $E_C \approx 0.326 \Delta_0$ energy of superconducting state (at the position of local minima over $\Delta$ on the orange curve) becomes equal to the energy of Coulomb glass state: this is the location of first-order transition.  In terms of actual magnitude of minimal superconducting gap $|\Delta|$ before the transition, its location is given by $E_C \approx 0.564 |\Delta|$ within our mean-field approximation.

It is important to note that near the critical value of $E_C/\Delta_0$, the energy of the superconducting state appears positive compared to the normal (unpaired) state. This might seem counterintuitive, as superconductivity typically lowers the energy of a system. 
In our case, the unusual behavior arises from the anomalous first term on the R.H.S. of Eq.~\eqref{eq:SC:FS}. This term describes the increase in Coulomb energy due to the formation of a superconducting gap $\Delta$ and related decrease of the screening response. In terms of pseudospin representation it corresponds to the suppression of susceptibility to the Z component of conjugated field due to ordering in XY plane.

Throughout this Section we considered replica-symmetric solution only. The reason is that the instability condition with respect to 
replica-symmetry breaking is never observed within superconducting state (at least for relatively small $E_C$), as demonstrated in Appendix~\ref{sec:appendix:SCStability}.
On the conceptual level, this kind of behavior seems to be related with the specific nature of interaction leading (when it is strong enough) 
to a glassy state. In our case it is Coulomb interaction and superconducting pairing modifies it  via the suppression of the susceptibility to the Z component of conjugated field.  We could imagine that in the case of the infinite-range random ZZ interaction  the situation might be different, leading to some co-existence region  of XY-plane ordering and glassiness, like it was discussed in Ref.~\cite{Markus2012}; this issue needs further investigation.

\section{Phase diagram}
\label{sec:PhaseDiag}

\begin{figure}
    \includegraphics[width=0.6\textwidth]{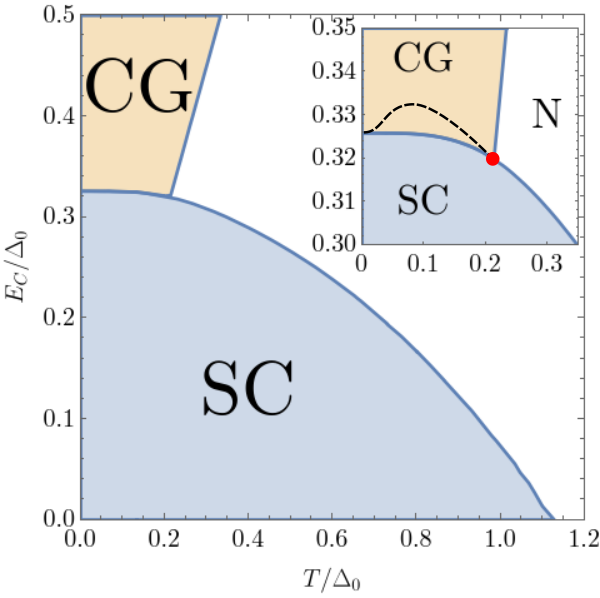}
    \centering
    \caption{Mean-field phase diagram on $(T/\Delta_0, E_C / \Delta_0)$ plane, with blue superconducting (SC) region, orange Coulomb glass (CG) region and white normal (N) insulating region. Inset: zoomed low-temperature region. Dashed line: suggested phase boundary for a more physical model with inhomogeneous order parameter $\Delta(\boldsymbol{r})$, see discussion in Sec. \ref{sec:Discussion}.}
    \label{fig:phasediag}
\end{figure}

With a complete mean-field description of both the superconducting and Coulomb glass phases, let us now analyze the mean-field phase diagram presented in Figure~\ref{fig:phasediag}. This diagram depicts three distinct phases: Coulomb glass insulating phase (CG): described in detail in Section~\ref{sec:CG}, Superconducting phase (SC): described in Section~\ref{sec:SC}, and Normal insulating state (N).

The boundary between the normal and Coulomb glass phases is the Almeida-Thouless line (Eq.~\eqref{eq:G:Temp}). This transition is continuous and characterized by the Parisi infinite-order replica symmetry breaking scheme. The transition between the normal and superconducting states occurs when a global minimum appears in the superconducting free energy (including Coulomb corrections) as a function of the absolute value of the superconducting gap $|\Delta|$. In the absence of Coulomb interactions, the transition would occur at a critical temperature ($T_S$) that is twice the standard BCS value due to the absence of single-particle excitations in our model. This translates to $T_{S}/\Delta_{0}=2e^{\gamma}/\pi\approx1.134$.

The transition between the superconducting and Coulomb glass phases is first-order, characterized by simultaneous jumps in both the glass order parameter $Q(x)$ and the superconducting order parameter $\Delta$. This transition occurs when the free energies of the two states, given by Eqs. (\ref{eq:G:FreeEnergy}) and (\ref{eq:SC:FreeEnergy}), become equal. 
Within mean-field approximation, at low temperatures, the dependence of the superconducting state's free energy on temperature has an activation form: $\mathcal{F}_{S}(T \to 0)-\mathcal{F}_{S}(T)\sim\exp(-|\Delta|/T)$. However, this contribution is negligible due to the presence of a hard superconducting gap $|\Delta|$ in the excitation spectrum (not to be confused with the large single-particle gap $\Delta_P$). In contrast, the temperature dependence of the Coulomb glass free energy can be derived from its entropy (Eq. (\ref{eq:G:Ent:LowT})) and is given by: ${\cal F}_{G}(T\to0)-{\cal F}_{G}(T)\sim \nu_0 E_{C}^2\,(T/E_{C})^{4}$.
This temperature dependence of the free energies determines the shape of the continuous line in the inset of Fig.~\ref{fig:phasediag} that separates the superconducting phase from the Coulomb glass phase. The line favors the Coulomb glass state at higher temperatures due to its stronger temperature dependence.

\section{Discussion and Conclusions}
\label{sec:Discussion}

Our results demonstrate that the pseudo-gap superconducting state undergoes an abrupt transition to an insulating state at a critical minimum value of the superconducting gap $\Delta_{\text{min}}$. This means that with a slightly higher level of disorder, $\Delta$ vanishes completely, along with the superfluid stiffness $\Theta$. This phase transition occurs when $\Delta_{\text{min}}$ becomes comparable in magnitude to the Coulomb gap $E_C$ characterizing the insulating ground state on the other side of the transition.
We focused on the bulk (3D) scenario because the films studied in~\cite{Exp-1st-order} are thick ($d \gg \xi_0$), where $E_C$ is determined by the density of states at the Fermi level $\nu_0$ and the macroscopic dielectric constant $\varepsilon$ in the insulating state. Consequently, the minimum gap $\Delta_{\text{min}}$, the minimum superfluid stiffness $\Theta_{\text{min}}$, and the maximum kinetic inductance $L_K^{\text{max}} \propto 1/\Theta_{\text{min}}$ are all ultimately determined by $\nu_0$ and $\varepsilon$.

While a general equation relating the kinetic inductance $L_K$ to the gap value $\Delta$ isn't available for pseudo-gap superconductors, we can leverage an empirical relation from Ref.~\cite{Exp-1st-order} to estimate the maximum achievable kinetic inductance:
\begin{equation}
L_K \approx \mathcal{A}\frac{c^2\hbar R_\Box}{\pi \Delta}
\label{Lk-1}
\end{equation}
Here, $\mathcal{A}$ is a numerical factor equal to 1 within the dirty BCS theory, but data suggests it can be around 3-5 for pseudo-gap superconductors. Additionally, $R_\Box$ represents the Drude resistance per square in the normal state, measured well above the transition temperature. 
Using Eq. (\ref{Lk-1}) and assuming $\Delta \approx E_C$ at the transition point, we can derive a formula for the maximum kinetic inductance ($L_K^{\text{max}}$):
\begin{equation}
L_K^{\text{max}} \approx \frac{3\mathcal{A}}{g \sqrt{\tilde\nu_0}}\left(10^{-3} \varepsilon\right)^{3/2} \mathrm{nH}
\label{Lk-2}
\end{equation}
In this equation, $g=2\pi\hbar/4e^2R_\Box$ represents the dimensionless film conductance, and $\tilde\nu_0$ is the reduced density of states defined by the relation $\nu_0 = \tilde\nu_0\cdot 10^{33} \mathrm{cm^{-3}erg^{-1}}$. 
Refs.~\cite{Exp-1st-order} reports a maximum experimentally measured kinetic inductance close to 17 nH, which aligns reasonably well with Eq. (\ref{Lk-2}). This is because both $g$ and $\tilde\nu_0$ are of order of unity, while $\epsilon \approx 1500$ based on extrapolations from Ref.~\cite{Ebensperger}. It is important to acknowledge that within this simplified theoretical framework, factors of order unity might not be entirely reliable due to the approximations made (discussed further below).

Following a similar approach for thin films with thickness $d \ll \xi_0$, we can combine Eq. (\ref{Ec2d}) with Eq. (\ref{Lk-1}) and the estimate $\Delta \approx E_C^{(\text{2D})}$ to arrive at the result for the maximum kinetic inductance in a 2D system:
\begin{equation}
    L_K^{\text{max},(\text{2D})} \approx \frac{\mathcal{A}}{g}\, d\, (10^{-3}\varepsilon ) \,  \mathrm{nH}
    \label{Lk2d}
\end{equation}
Here, the film thickness ($d$) is measured in nanometers. This estimate (Eq. (\ref{Lk2d})) might be particularly relevant for very thin ($\sim 2~\text{nm}$) NbN films, which exhibit characteristics of pseudo-gap superconductivity. Equations (\ref{Lk-2}) and (\ref{Lk2d}) represent key practical implications of our theory, offering valuable guidance for the design of ``super-inductors''.

While the location of the zero-temperature transition between the competing phases seems straightforward ($\Delta \approx E_C$), determining the exact shape of the transition line in the  $\left(T / \Delta_0, E_C / \Delta_0\right)$ plane, even at low temperatures ($T\ll \Delta$), becomes more challenging. 
At finite temperatures, the phase transition boundary is determined by the equality of free energies: $\mathcal{F}_S(T) = \mathcal{F}_G(T)$. The sign of the derivative $d(\mathcal{F}_S(T)-\mathcal{F}_G(T))/dT = T(S_G(T)-S_S(T))$ (where $S_{S,G}(T)$ are the entropy of the superconducting and glassy states, respectively) indicates the direction of the slope $dE_C^*(T)/dT$ for the transition line in Fig. \ref{fig:phasediag}. 
Our results in Section~\ref{sec:CG} demonstrate that the glass entropy $S_G$ scales as $T^3$, leading to a higher entropy compared to the exponentially small entropy of the superconducting state in the $T\to 0$ limit. However, this contradicts the expected behavior in real, strongly disordered superconductors, as we will discuss further below.

It is important to discuss the approximations made in our model and their potential impact. The primary approximation is neglecting spatial variations of the order parameter $\Delta(\boldsymbol{r})$, known to be significant in pseudo-gap superconductors~\cite{NaturePhys2011,Anton1}. This undoubtedly limits the accuracy of our theory (in terms of the transition point and minimum $\Theta$ value) by an unknown factor of order one. However, it shouldn't affect the overall observation of a sharp drop in $\Theta$, as this arises from the competition between ground states with fundamentally different order parameters. 
However, accounting for order parameter inhomogeneity does influence the shape of the phase diagram in Fig.~\ref{fig:phasediag}. According to Ref.~\cite{Anton2}, temperature variations of the superfluid stiffness follow a power law: $\Theta(0)-\Theta(T) \propto T^\beta $, with a non-universal exponent $\beta \approx 1.6 - 2.5$. This power law stems from the existence of very low-energy excitations due to the broad distribution of order parameter magnitudes in space. These same excitations lead to an entropy contribution $S_S\propto T^\beta$ that becomes \textit{larger} than the glass entropy $S_G \propto T^3$ at low temperatures. Consequently, the actual position of the critical line $dE_C^*(T)/dT$ is expected to behave as shown by the black dashed line in the inset of Fig.~\ref{fig:phasediag}, which aligns better with experimental data~\cite{Exp-1st-order}.

Since this is a first-order phase transition, the potential existence of metastable states becomes relevant. In the experimental protocol of Refs.~\cite{Exp-1st-order}, no metastable behavior was observed as the samples with varying disorder levels were cooled down and measured at low temperatures. However, microscopic disorder in the spatial distribution of $\Delta(\boldsymbol{r})$ could lead to larger-scale (mesoscopic) inhomogeneities, where superconducting and insulating states form a kind of domain structure. The typical spatial scale of these domains might be substantial. We speculate that this phenomenon could explain the striking results of Ref.~\cite{Kowal-Ovadyahu}, where significant size-dependent transport properties were observed in InO$_x$ films with sizes up to tenths of microns.

Investigating the nature of metastable states near this first-order SIT would benefit from experimental techniques that allow control of the transition with a continuously tunable parameter, such as a magnetic field. However, it is important to remember that in the presence of a magnetic field, there is no straightforward relationship between the superfluid stiffness $\Theta$ measured in experiments like~\cite{Exp-1st-order} and the superconducting gap $\Delta$ that competes with the Coulomb gap $E_C$. Therefore, understanding the behavior of the magnetic-field-driven SIT requires dedicated studies of the  interplay between all these factors.

Our analysis hinges on estimating the Coulomb interaction strength at moderate (Section \ref{sec:CoulombPhononCompetition}) and long scales (Section \ref{sec:CoulombRepulsion}). The first subsection demonstrates how the very existence of superconductivity in an amorphous material with high resistivity and low electron density can be understood through the fractal nature of electron wavefunctions. The second subsection shows that while Coulomb repulsion is \textit{relatively} suppressed at short distances, it re-emerges at longer scales and eventually disrupts superconductivity under extremely strong disorder.

On a more technical note, our findings include a detailed study of the thermodynamic properties of the Coulomb glass state, which was absent in previous works~\cite{MullerIoffe,MullerPankov}.

\section*{Acknowledgments}

We are grateful to Denis Basko, Anton Khvalyuk, Vladimir Kravtsov and Benjamin Sac\'ep\'e for numerous helpful discussions. 

\paragraph{Funding information}
I.P. acknowledges support by the Deutsche Forschungsgemeinschaft (DFG) via the grant MI 658/14-1.

\begin{appendix}

\section{Derivation of the mean field free energy functional}
\label{sec:appendix:MeanField}
We start our analytical treatment of the model \eqref{H0} by decoupling the interaction via two Hubbard-Stratonovich fields, one having the meaning of the complex order parameter $\Delta$ and other being the plasmonic field $\phi$:
\begin{multline}
e^{-\beta F}=\int{\cal D}\phi{\cal D}\Delta\exp\left(-\int_{0}^{\beta}d\tau\left[\bar{\Delta}_{i}(\tau)\hat{J}_{ij}^{-1}\Delta_{j}(\tau)+\frac{1}{2}\phi_{i}(\tau)\hat{U}_{ij}^{-1}\phi_{j}(\tau)\right]\right)\\
\times\prod_{i}{\rm Tr}{\cal T}_{\tau}\exp\left(-\int_{0}^{\beta}d\tau\hat{H}_{\text{loc}}(\tau)\right)
\end{multline}
with $\beta = 1/T$ and 
\begin{equation}
\hat{H}_{\text{loc}}(\tau)=2(\xi_{i}+i\phi_{i}(\tau))\hat{S}_{i}^{z}(\tau) -\left(\Delta_{i}(\tau)\hat{S}_{i}^{-}(\tau)+h.c.\right)
\end{equation}

The averaging over the distribution of random fields $\xi_i$ is performed utilizing the standard replica trick. Fields $\Delta_i$, $\phi_i$ and spin degrees of freedom $\hat{S}_i$ acquire additional replica index $a = 1, \dots, n$, and the quenched disorder leads to the mixing between replicas:
\begin{equation}
    \overline{\exp\left(-2\int_{0}^{\beta}d\tau\xi_{i}\sum_{a=1}^{n}\hat{S}_{i,a}^{z}(\tau)\right)}\\
    =\exp\left(2W^{2}\int_{0}^{\beta}d\tau d\tau^{\prime}\hat{S}_{i,a}^{z}(\tau){\cal I}_{ab}\hat{S}_{i,b}^{z}(\tau^{\prime})\right)
\end{equation}
with all matrix elements of matrix ${\cal I}_{ab}$ are equal to unity.

As was explained in the main text, we will treat superconductivity in the mean-field approximation and neglect spatial and temporal fluctuations of the superconducting order parameter $\Delta_{i, a}(\tau) = \Delta = \const$. 

Following Refs. \cite{MullerIoffe,MullerPankov}, we identify self-consistently RPA-screened ``cactus'' diagrams as giving the leading contribution to the free energy in the limit of large disorder $W \gg \Delta, E_C$ and as being responsible for the freezing transition. Such diagrams can be resummed self-consistently by introducing the auxillary local Coulomb energy matrix $G_{i}^{ab}(\tau,\tau^{\prime})\equiv U_{ii}\delta^{ab}\delta(\tau-\tau^{\prime})-\left\langle\phi_{i}^{a}(\tau)\phi_{i}^{b}(\tau^{\prime})\right\rangle$ (with the first term acting as a counter-term cancelling unphysical ``self-interaction'') and the Lagrange multiplier $Q_{i}^{ab}(\tau,\tau^{\prime})$ which imposes this condition:
\begin{multline}
    -\beta nF_{L}[\hat{G},\hat{Q},\phi]=-\frac{1}{2}\sum_{i,a,b}\int_{0}^{\beta}d\tau d\tau^{\prime}\left(G_{i}^{ab}(\tau,\tau^{\prime})Q_{i}^{ab}(\tau,\tau^{\prime})+\phi_{i}^{a}(\tau)Q_{i}^{ab}(\tau,\tau^{\prime})\phi_{i}^{b}(\tau^{\prime})\right)
    \\
    -\frac{1}{2}\int_{0}^{\beta}d\tau\sum_{i,a}U_{ii}Q_{i}^{aa}(\tau,\tau)
\end{multline}
and, as we can see show below, has the meaning of local polarizability and spin glass order parameter. Resummation of the cactus diagram then equivalent~\cite{Poboiko2020} 
to the following replacement in the action describing local spin degrees of freedom:
\begin{equation}
    \exp\left(-2i\int_{0}^{\beta}d\tau\phi_{i,a}(\tau)\hat{S}_{i,a}^{z}(\tau)\right)
    \mapsto\exp\left(2\int_{0}^{\beta}d\tau d\tau^{\prime}\hat{S}_{i,a}^{z}(\tau)G_{i}^{ab}(\tau,\tau^{\prime})\hat{S}_{i,b}^{z}(\tau^{\prime})\right)
\end{equation}

The remaining Gaussian integration over $\phi$-fields yields:
\begin{equation}
    \int{\cal D}\phi\exp\left(-\frac{1}{2}\phi(\hat{U}^{-1}+\hat{Q})\phi\right)=\exp\left(-\frac{1}{2}\Tr\ln(1+\hat{U}\hat{Q})\right)
\end{equation}
which brings us to the final form of the mean field free energy functional, which now has to be minimized w.r.t. $G$, $Q$ and $\Delta$:
\begin{multline}
\label{eq:app:F}
-\beta nF[\hat{G},\hat{Q},\Delta]=-\frac{1}{2}\Tr\left(\hat{G}\hat{Q}-\hat{U}\hat{Q}+\ln(1+\hat{U}\hat{Q})\right)\\
+\sum_{i}\left(-\beta nP_{0}|\Delta|^{2}/\lambda+\ln\Tr_{S}{\cal T}_{\tau}\exp\left(-A_{\text{loc}}[\hat{G}_{i},\hat{S}_{i},\Delta_{i}]\right)\right),
\end{multline}
where we have introduced dimensionless superconducting coupling constant $\lambda = P_0 \sum_i J_{ij}$, the spectrum density at the Fermi level $P_0 \equiv P_0(\xi = 0)$, and local in real space spin action given by Eq. \eqref{eq:LocalSpinActionQuantum} in the main text. The traces are calculated over the Matsubara imaginary time space, replica space and over the localized single-electron orbitals.

Within the mean-field approximation, we assume matrices $\hat{G}$ and $\hat{Q}$ to be spatially homogeneous $G_{i}^{ab}(\tau,\tau^{\prime})=G^{ab}(\tau,\tau^{\prime})$. Assuming that localized states are distributed homogeneously in the real space with large concentration $n_{0}$, we replace summation over the sites by the integration over real space as $\sum_{i}f_{i}\mapsto n_{0}\int d\boldsymbol{r}f(\boldsymbol{r})$, leading to appearance of factors $n_0$ in the Eq. \eqref{eq:app:F}. This allows us to calculate the trace over real space and we arrive at the following expression for the free energy \emph{density} $\mathcal{F} \equiv F / V \equiv \mathcal{F}_C[\hat{G},\hat{Q}] + \mathcal{F}_\text{loc}[\Delta,\hat{G}]$, with:
\begin{equation}
{\cal F}_{C}[\hat{G},\hat{Q}]=\frac{T}{2n}{\rm Tr}\left(n_{0}\hat{G}\hat{Q}-\Phi(\hat{Q})\right),\quad \Phi(Q)=\int(d^3\boldsymbol{k})\left(n_{0}U_{\boldsymbol{k}}Q-\ln(1+n_{0}U_{\boldsymbol{k}}Q)\right)
\end{equation}
\begin{equation}
{\cal F}_{\text{loc}}[\hat{G},\Delta]=\frac{\nu_{0}|\Delta|^{2}}{\lambda}-\frac{T}{n}n_{0}\ln{\rm Tr}{\cal T}_{\tau}\exp\left(-\hat{A}_{\text{loc}}[\hat{G},\Delta]\right),
\end{equation}
where trace in $\mathcal{F}_C$ now taken over Matsubara and replica spaces only, and $\nu_0 \equiv n_0 P_0$ is the density of states (per unit volume).

\section{Parisi ansatz for ultrametric matrices}
\label{sec:appendix:RSB}
To describe the glass phase, we will utilize standard Parisi ultrametric replica symmetry breaking ansatz for arbitrary matrix in a replica space $\hat{M}$ as follows:
\begin{equation}
M^{ab}=M_{0}{\cal I}^{ab}+\sum_{k=1}^{R+1}(M_{k}-M_{k-1}){\cal I}_{k}^{ab},
\end{equation}
with $R$ being number of replica symmetry breaking steps, and matrices ${\cal I}_{k}^{ab}=\delta_{[a/m_{k}],[b/m_{k}]}$, where $[...]$ denotes the integer part, $\delta$ being Kronecker delta-symbol and $n = m_0 > m_1 > \dots > m_{R+1} = 1$ being sizes of corresponding blocks in the replica symmetry breaking scheme. In the limit $R \to \infty$, which corresponds to the continuous RSB, the replica structure of these matrices can then be encoded in a single function defined on $x \in [0,1]$ as follows:
\begin{equation}
    M(x) = M_0 + \sum_{k=1}^{R+1}(M_k - M_{k-1}) \theta(x - m_k),
\end{equation}
with $\theta(x)$ being Heaviside theta-function.

We will also utilize related representation, which is called ``Replica Fourier Transform'' of a matrix $\hat{M}$. The representation, and its inverse, are defined as follows:
\begin{align}
    \widetilde{M}_{k}&=\sum_{i=k}^{R+1}m_{i}(M_{i}-M_{i-1}),\quad k = 1, \dots, R+1,\\
    M_{k}&=M_{0}+\sum_{i=1}^{k}(\widetilde{M}_{i}-\widetilde{M}_{i+1})/m_{i}.
\end{align}
However, from this definition it follows that $\widetilde{M}_{0}-\widetilde{M}_{1}=nM_{0}$, which vanishes in the replica limit; for this reason, the component $M_0 = M(0)$ should be kept as a separate parameter in the RFT parametrization. 

The continuous RFT defined as:
\begin{equation}
\widetilde{M}(x)\equiv\widetilde{M}_{R+1}\theta(1-x)+\sum_{k=0}^{R}(\widetilde{M}_{k}-\widetilde{M}_{k+1})\theta(m_{k}-x),
\end{equation}
and has the property 
\begin{equation}
\widetilde{M}^{\prime}(x)=-xM^{\prime}(x)\Rightarrow \widetilde{M}(x)=\int_{x}^{1_{+}}dx\,xM^{\prime}(x)=M(1_{+})-xM(x)-\int_{x}^{1}M(y)dy
\end{equation}
where $M(1_+)$ corresponds to diagonal elements $M_{R+1}$, as function $M(x)$ is, strictly speaking, discontinuous at $x = 1$.

This representation emerges naturally in a Replica Fourier basis, which diagonalizes arbitrary ultrametric matrix; quantities $\widetilde{M}_{k}$ denote eigenvalues of matrix $M$ with degeneracies ${\rm dim}_k = n/m_k - n/m_{k-1}$. This dictates following very useful properties of RFT:
\begin{itemize}
    \item If $\hat{C} = \hat{A} \hat{B}$, then 
\begin{equation}
    \widetilde{C}(x)=\widetilde{A}(x)\widetilde{B}(x),\quad C(0)=\widetilde{A}(0)B(0)+A(0)\widetilde{B}(0)
\end{equation}
    \item If $\hat{B} = f(\hat{A})$, then:
\begin{equation}
    \widetilde{B}(x)=f(\widetilde{A}(x)),\quad B(0)=A(0)f^{\prime}(\widetilde{A}(0))
\end{equation}
    \item Finally, the trace of an ultrametric matrix can be calculated as:
\begin{equation}
\frac{1}{n}\tr\hat{A}=A(1_{+})=A(0)+\widetilde{A}(0)+\int_{0}^{1}dx\left(1-\frac{1}{x}\right)\widetilde{A}^{\prime}(x)
\end{equation}
\end{itemize}

\section{Parisi free energy}
\label{sec:appendix:Parisi}
As discussed in the beginning of the Sec. \ref{sec:CG}, the full free energy of the classical Coulomb glass consists of two parts:
\begin{equation}
\label{eq:app:Ftot}
{\cal F}[\hat{G},\hat{Q}]={\cal F}_{\text{loc}}[\hat{G}]+{\cal F}_C[\hat{G},\hat{Q}],
\end{equation}
with the ``Coulomb'' contribution
\begin{equation}
\label{eq:app:Fc}
{\cal F}_{C}[\hat{G},\hat{Q}]=\frac{T}{2n}{\rm Tr}\left(n_{0}\beta^{2}\hat{G}\hat{Q}-\Phi(\beta\hat{Q})\right),
\end{equation}
and the ``local'' contribution 
\begin{equation}
\label{eq:app:Floc}
{\cal F}_{\text{loc}}[\hat{G}]=-\frac{T}{n}n_{0}\ln{\rm Tr}\exp\left(-2\beta^{2}S_{a}\left(W^{2}{\cal I}^{ab}+G^{ab}\right)S_{b}\right),
\end{equation}
In this Appendix, we will present its calculation for arbitrary form of the function $\Phi(\hat{Q})$. 

\subsection{Parisi scheme}
The ``local'' contribution can be expressed in a standard way via the Parisi free energy $f(x,\xi) = - T \ln Z(x,\xi)$ as follows:
\begin{equation}
{\cal F}_{\text{loc}}[G]=n_{0}\int_{-\infty}^{+\infty} d\xi_{0}\widetilde{P}_{0}(\xi_{0})f(x=0,\xi_0)-\frac{\beta}{2}n_{0}\widetilde{G}(0),
\end{equation}
with $\widetilde{G}(x)$ denoting the Replica Fourier Transform of matrix $\hat{G}$ (see Appendix \ref{sec:appendix:RSB}), and the ``renormalized'' distribution function
\begin{equation}
\widetilde{P}_{0}(\xi)=\frac{\exp\left(-\xi^{2}/2\widetilde{W}^{2}\right)}{\sqrt{2\pi\widetilde{W}}},\quad\widetilde{W}^{2}\equiv W^{2}+G(0).
\end{equation}
In the limit $W \to 0$ this reduces to the standard Parisi scheme e.g. for Sherrington-Kirkpatrick model, while in the present calculation we will be interested in the opposite limit $W \to \infty$. The function $f(x,\xi)$ satisfies the Parisi partial differential equation
\begin{equation}
\partial_{x}f=-\frac{1}{2}G^{\prime}(x)\left(\partial_{\xi}^{2} f+\beta x\left(1-(\partial_{\xi}f)^{2}\right)\right),\quad f(x=1,\xi)=-T\ln2\cosh\beta\xi,
\end{equation}
which can be equivalently rewritten via the magnetization $m(x,\xi)\equiv-\partial_{\xi}f(x,\xi)$:
\begin{equation}
\label{eq:app:parisim}
-\partial_{x}m=\frac{1}{2}G^{\prime}(x)\left(\partial_{\xi}^{2}m+\beta x\partial_{\xi}(m^{2})\right),\quad m(x=1,\xi)=\tanh\beta\xi.
\end{equation}

Within the saddle point approximation, the $Q$-matrix is determined by:
\begin{equation}
\label{eq:app:QFlocDeriv}
Q_{ab}=4\left\langle S_{a}S_{b}\right\rangle_G =-\frac{2T}{n_{0}}\frac{\delta (n{\cal F}_{\text{loc}}[G])}{\delta G_{ab}},
\end{equation}
with the average calculated w.r.t. exponential weight defined by Eq. \eqref{eq:app:Floc}.
It can be expressed in terms of auxillary function $P(x,\xi)$ having the meaning of distribution function of frozen fields:
\begin{equation}
\label{eq:app:Q}
\widetilde{Q}(x)=T\int_{-\infty}^{+\infty} d\xi P(x,\xi)\partial_{\xi}m(x,\xi),\quad Q(x)=\int_{-\infty}^{+\infty} d\xi P(x,\xi)m^{2}(x,\xi),
\end{equation}
and which satisfies following equation:
\begin{equation}
\label{eq:app:parisip}
\partial_{x}P=\frac{1}{2}G^{\prime}(x)\left(\partial_{\xi}^{2}P-2\beta x\partial_{\xi}(mP)\right),\quad P(x=0,\xi) = \widetilde{P}_0(\xi).
\end{equation}
Finally, the extremum of the free energy w.r.t $\hat{Q}$ yields:
\begin{equation}
\label{eq:app:coupling}
\beta n_0 \hat{G}=\Phi^{\prime}(\beta\hat{Q}),\quad \Rightarrow \quad n_0 \widetilde{G}(x)=T\Phi^{\prime}(\beta\widetilde{Q}(x)),\quad  n_0 G(0)=Q(0)\Phi^{\prime\prime}(\beta\widetilde{Q}(0)).
\end{equation}
This closes the self-consistency scheme $m(x,\xi) \mapsto P(x,\xi) \mapsto Q(x) \mapsto G(x) \mapsto m(x,\xi)$. 

Differentiating \eqref{eq:app:Q} w.r.t. temperature, we arrive at:
\begin{equation}
\widetilde{Q}^{\prime}(x)=\widetilde{G}^{\prime}(x)\int_{-\infty}^{+\infty} d\xi P(x,\xi)(\partial_{\xi}m(x,\xi))^{2}
\end{equation}
from which follows the marginal stability criterion:
\begin{equation}
\label{eq:app:MarginalStability}
1=n_{0}^{-1}\Phi^{\prime\prime}(\beta\widetilde{Q}(x))\int_{-\infty}^{+\infty} d\xi P(x,\xi)(\partial_{\xi}m(x,\xi))^{2},
\end{equation}
which has to be fullfilled as long as $\widetilde{Q}^\prime(x) \neq 0$.

\subsection{Free energy, internal energy and entropy}
The Coulomb contribution can be calculated utilizing the RFT, and yields:
\begin{multline}
    {\cal F}_{C}[G,Q]=\frac{1}{2}\Bigg(\beta n_{0}\widetilde{G}(1)\widetilde{Q}(1)-T\Phi(\beta\widetilde{Q}(1))+\beta n_{0}G(0)\widetilde{Q}(0)\\
+Q(0)\left(\beta n_{0}\widetilde{G}(0)-\Phi^{\prime}(\beta\widetilde{Q}(0))\right)\Bigg)
-\frac{1}{2}\int_{0}^{1}\frac{dx}{x}\frac{d}{dx}\left(\beta n_{0}\widetilde{G}(x)\widetilde{Q}(x)-T\Phi(\beta\widetilde{Q}(x))\right),
\end{multline}
and can be further simplified utilizing Eq. \eqref{eq:app:coupling}:
\begin{equation}
{\cal F}_{C}=\frac{1}{2}\left(-T\Phi(\beta\widetilde{Q}(1))+\beta n_{0}\left(\widetilde{G}(1)\widetilde{Q}(1)+G(0)\widetilde{Q}(0)+\int_{0}^{1}dxG^{\prime}(x)\widetilde{Q}(x)\right)\right).
\end{equation}
Thus the full free energy reads:
{\small
\begin{equation}
\label{eq:app:FGlass:result}
    {\cal F}=n_{0}f(0)+\frac{1}{2}\left(-T\Phi(\beta\widetilde{Q}(1))+\beta n_{0}\left(\widetilde{G}(1)\widetilde{Q}(1)-\widetilde{G}(0)+G(0)\widetilde{Q}(0)+\int_{0}^{1}dxG^{\prime}(x)\widetilde{Q}(x)\right)\right),
\end{equation}
}
where we have introduced 
\begin{equation}
f(x)\equiv \int_{-\infty}^{+\infty} d\xi\,P(x,\xi)f(x,\xi)
\end{equation}

This partially averaged self-energy satisfies the following equation, which can be derived directly from the Parisi equations:
\begin{equation}
f^{\prime}(x)=\frac{1}{2}\beta \widetilde{G}^{\prime}(x)(1+Q(x)),
\end{equation}
which allows us to express the full free energy equivalently via $f(1)$:
\begin{multline}
{\cal F}=n_0 f(1)+\frac{1}{2}\Bigg(-T\Phi(\beta\widetilde{Q}(1))\\
+\beta n_{0}\left(G(0)\widetilde{Q}(0)-\widetilde{G}(1)Q(1) + \int_{0}^{1}dx\left(G^{\prime}(x)\widetilde{Q}(x)-\widetilde{G}^{\prime}(x)Q(x)\right)\right)\Bigg)
\end{multline}

The internal energy $\mathcal{E} = \partial_\beta (\beta \mathcal{F})$ can be calculated directly from Eqs. (\ref{eq:app:Ftot}-\ref{eq:app:Floc}) by differentiating it w.r.t. $\beta$ and noting that one only has to take into account the explicit $\beta$-dependence as the free energy is already minimized w.r.t. $G$ and $Q$. As consequence, we obtain:
\begin{equation}
\label{eq:app:E}
{\cal E}=\partial_{\beta}(\beta{\cal F})=-\beta n_0 W^{2}\frac{1}{n}{\rm Tr}(\hat{Q}\hat{{\cal I}})-\frac{1}{2n}{\rm Tr}\partial_{\beta}\Phi(\beta\hat{Q})=-\beta n_{0} W^{2}\widetilde{Q}(0)-\frac{1}{2n}{\rm Tr}(\hat{Q}\Phi^{\prime}(\beta\hat{Q}))
\end{equation}

Utilizing Eq. \eqref{eq:app:Q} and the fact that $P(x=0,\xi)$ is a Gaussian distribution of width $\widetilde{W}$, we can integrate by parts and obtain:
\begin{equation}
\widetilde{Q}(0)=T\int_{-\infty}^{+\infty} d\xi\widetilde{P}_{0}(\xi)\partial_{\xi}m(0,\xi)=\frac{T}{\widetilde{W}^{2}}\int_{-\infty}^{+\infty} d\xi\widetilde{P}_{0}(\xi)\xi m(0,\xi).
\end{equation}
Reexpressing $W$ via $\widetilde{W}$ in Eq. \eqref{eq:app:E}, and calculating the trace, we arrive at:
\begin{equation}
{\cal E}=E(0)+\frac{\beta n_{0}}{2}\left(-\widetilde{G}(1)+G(0)\widetilde{Q}(0)+\int_{0}^{1}dx\left(\widetilde{G}^{\prime}(x)Q(x)-G^{\prime}(x)\widetilde{Q}(x)\right)\right),
\end{equation}
where we have introduced:
\begin{equation}
E(x)\equiv-\int_{-\infty}^{+\infty} d\xi P(x,\xi)\,\xi m(x,\xi)
\end{equation}
Utilizing Parisi equations, we can derive the following:
\begin{equation}
E^{\prime}(x)=-\beta G^{\prime}(x)\left(\widetilde{Q}(x)+xQ(x)\right),
\end{equation}
which allows us to express the internal energy via $E(1)$ as follows:
\begin{equation}
{\cal E}=n_0 E(1)+\frac{\beta n_{0}}{2}\left(-\widetilde{G}(1)+G(0)\widetilde{Q}(0)+\int_{0}^{1}dx\left(G^{\prime}(x)\widetilde{Q}(x)-\widetilde{G}^{\prime}(x)Q(x)\right)\right)
\end{equation}

As consequence, for the entropy $S = \beta(\mathcal{E} - \mathcal{F})$ we obtain two equivalent representations:
\begin{multline}
\label{eq:app:Entropy}
S=\beta n_{0}(E(0)-f(0))+\frac{1}{2}\Bigg(-\beta\widetilde{Q}(1)\Phi^{\prime}(\beta\widetilde{Q}(1))+\Phi(\widetilde{Q}(1))\\
+\beta^{2}n_{0}\left(\widetilde{G}(0)-\widetilde{G}(1)+\int_{0}^{1}dx\left(\widetilde{G}^{\prime}(x)Q(x)-2G^{\prime}(x)\widetilde{Q}(x)\right)\right)\Bigg)\\
=\beta n_{0}(E(1)-f(1))-\frac{1}{2}\left[\beta\widetilde{Q}(1)\Phi^{\prime}(\beta\widetilde{Q}(1))-\Phi(\widetilde{Q}(1))\right]
\end{multline}
The last equation is the most convenient one. The explicit integral representation for the first term reads:
\begin{equation}
\beta n_{0}(E(1)-f(1))=n_{0}\int_{-\infty}^{+\infty} d\xi\,P(1,\xi)\,(\ln2\cosh\beta\xi-\beta \xi\tanh\beta\xi)
\end{equation}

\subsection{\texorpdfstring{$W \to \infty$}{W to infinity} limit}
The scheme can be greatly simplified under the assumption of large bandwidth, which here allows one to neglect the energy-dependence of the bare density of states and approximate it with a constant $P_{0}(\xi)=\exp\left(-\xi^{2}/2W^{2}\right)/\sqrt{2\pi}W\approx P_{0}\equiv(\sqrt{2\pi}W)^{-1}$. One just has to carefully subtract the high-energy contributions which are, however, insensitive to the glass transition, in all the quantities.

We make following dimensionless substitutions:
\begin{equation}
\hat{Q}\equiv2P_{0}T\hat{q},\quad\hat{G}\equiv 2E_{C}^{2}\hat{g},\quad b \equiv \beta E_{C}   
\end{equation}
We also pick (at this point arbitrary) energy scale $E_C$ and also introduce:
\begin{equation}
y\equiv\xi/E_{C},\quad P(x,\xi)\equiv P_{0}p(x,y),\quad \Phi(\beta\hat{Q})=4\nu_{0}E_{C}\phi(\hat{q}).
\end{equation}
For the case of Coulomb glass, the dimensionless function has the very simple form $\phi(q) = 2 q^{3/2} / 3$. The relevant equations from the previous subsection, Eqs. (\ref{eq:app:parisim}-\ref{eq:app:MarginalStability}), then acquire the following form:
\begin{align}
    -\partial_{x}m&=g^{\prime}(x)\left(\partial_{y}^{2}m+bx\partial_{y}(m^{2})\right),\quad m(x=1,y)=\tanh by,\label{eq:app:dimensionless:m}\\
    \partial_{x}p&=g^{\prime}(x)\left(\partial_{y}^{2}p-2bx\partial_{y}(mp)\right),\quad p(x=0,y)=1,\label{eq:app:dimensionless:p}\\
    \widetilde{q}(x)&=\frac{1}{2}\int_{-\infty}^{+\infty} dy\,p(x,y)\partial_{y}m(x,y),\label{eq:app:dimensionless:qt}\\
    q(x)&=\frac{1}{2P_{0}T}-\frac{b}{2}\int_{-\infty}^{+\infty} dyp(x,y)(1-m^{2}(x,y)),\\
    \widetilde{g}(x)&\equiv\phi^{\prime}(q)/b\quad \Rightarrow\quad g^{\prime}(x)=\phi^{\prime\prime}(\widetilde{q}(x))q^{\prime}(x)/b,\label{eq:app:dimensionless:gp}\\
    1&=\phi^{\prime\prime}(\widetilde{q}(x))\int_{-\infty}^{+\infty} dyp(x,y)(\partial_{y}m(x,y))^{2},\label{eq:app:dimensionless:marginalstability}
\end{align}
which give Eqs. (\ref{eq:ParisiM}-\ref{eq:MarginalStability}) in the main text. 
The most important simplification here was replacement of the initial condition by a constant $P(0,\xi) = P_0$, which is justified by the parameter $W \gg T, E_C$. From this approximation it immediately follows that:
\begin{equation}
\widetilde{q}(0)=\frac{1}{2}\left(m(0,+\infty)-m(0,-\infty)\right)=1.
\end{equation}

Furthermore, the marginal stability criterion evaluated exactly at $T = T_G$ (below which $\widetilde{q}^\prime(x) \neq 0$ continuously appears), where $\widetilde{q}(x) = 1$, $m(x,y) = \tanh b_G y$, and $p(x,y) = 1$ gives the glass transition temperature:
\begin{equation}
T_{G}/E_{C}=\frac{4}{3}\phi^{\prime\prime}(1),
\end{equation}
which for the Coulomb glass problem yields Eq. \eqref{eq:G:Temp} from the main text.

We now switch to the calculation of the free energy and entropy. The first observation is that $G(0)$ only enters $\mathcal{F}_{\text{loc}}$ via small renormalization of large bandwidth $\widetilde{W}^2 = W^2 + G(0)$, thus its effect can be calculated via the Taylor expansion:
\begin{equation}
\label{eq:app:G0term}
{\cal F}_{\text{loc}}(G_{0})-{\cal F}_{\text{loc}}(G_{0}=0)\underset{W\to\infty}{\simeq}-\frac{1}{2}\beta n_0 G(0)\widetilde{Q}(0),
\end{equation}
where we have utilized Eq. \eqref{eq:app:QFlocDeriv} to calculate the derivative. After such expansion, we can safely replace $\widetilde{P}_0(\xi)$ by the bare distribution function $P_0(\xi)$ in the Parisi scheme. To calculate the free energy, we subtract from Eq. \eqref{eq:app:FGlass:result} the same contribution with $\widetilde{G}(x) = \widetilde{G}(0)$ and $\widetilde{Q}(x) = \widetilde{Q}(0)$, which is the free energy of the normal state; the main contribution to their difference then comes from the vicinity of the Fermi surface, justifying the constant density of states approximation. Taking also into account \eqref{eq:app:G0term}, and making dimensionless substitution, we arrive at:
\begin{multline}
{\cal F}_{G}-{\cal F}_{N}=\nu_{0}E_{C}\int_{-\infty}^{+\infty} dy\left(f(0,y)-f(1,y)\right)+2\nu_{0}E_{C}^{2}\int_{0}^{1}dxg^{\prime}(x)\widetilde{q}(x)\\
-2\nu_{0}E_{C}T\left(\phi^{\prime}(1)-\phi(1)-\widetilde{q}(1)\phi^{\prime}(\widetilde{q}(1))+\phi(\widetilde{q}(1))\right).
\end{multline}
The first integral can be conveniently expressed via $m(x,y) = -\partial_y f(x,y) / E_C$:
\begin{multline}
\int_{-\infty}^{\infty}dy\left(f(0,y)-f(1,y)\right)=E_{C}\lim_{\Lambda\to\infty}\int_{-\Lambda}^{\Lambda}dy\int_{-\Lambda}^{y}dy^{\prime}\left(m(1,y)-m(0,y)\right)\\
=E_{C}\lim_{\Lambda\to\infty}\int_{-\Lambda}^{\Lambda}dy\left(m(1,y)-m(0,y)\right)(\Lambda-y)=E_{C}\int_{-\infty}^{\infty}dy\,y\left(m(0,y)-m(1,y)\right),
\end{multline}
where we have utilized that $m(x,y)$ is an odd function of $y$. The last expression is nothing but the integral of the full derivative:
\begin{equation}
-\int_{0}^{1}dx\frac{d}{dx}(\widetilde{q}(x)\phi^{\prime}(\widetilde{q}(x))-\phi(\widetilde{q}(x)))=\beta E_{C}\int_{0}^{1}dx\,x g^\prime(x) \widetilde{q}(x),
\end{equation}
which yields Eq. \eqref{eq:G:FreeEnergy} from the main text. Finally, the entropy given by Eq. \eqref{eq:app:Entropy} is already defined by low-energy physics $\xi \sim \max(T, E_C)$, allowing us to make dimensionless substitution and arrive at:
\begin{equation}
S_{G}/\nu_{0}E_{C}=\int dyp(1,y)\left[\ln2\cosh by-by\tanh by\right]-2\left(\widetilde{q}(1)\phi^{\prime}(\widetilde{q}(1))-\phi(\widetilde{q}(1))\right),
\end{equation}
which yields Eq. \eqref{eq:G:Ent} from the main text.

\section{M\"uller-Pankov low-temperature scaling}
\label{sec:appendix:Pankov}
In the low-temperature regime $T \ll E_C$, the dimensionless Parisi equations formulated in the Appendix \ref{sec:appendix:Parisi} acquire universal scaling for $1 \gg x \gg T/E_C$ discussed in Ref. \cite{MullerPankov}. In the present Appendix we reproduce it for the generic power-law dependence
\begin{equation}
\phi(q)=\frac{n}{n+1}q^{1+1/n},
\end{equation}
with the 3D Coulomb glass corresponding to $n = 2$. 

We switch to variables $\tau \equiv b x \in [0, b]$ and $z \equiv b x y$, the Parisi equations (\ref{eq:app:dimensionless:m}-\ref{eq:app:dimensionless:marginalstability}) then read:
\begin{align}
-\left(z\partial_{z}m+\tau\partial_{\tau}m\right)&=\tau^{3}g^{\prime}(\tau)\left[\partial_{z}^{2}m+\partial_{z}(m^{2})\right],\quad m(\tau=b,z)=\tanh z\\
z\partial_{z}p+\tau\partial_{\tau}p&=\tau^{3}g^{\prime}(\tau)\left[\partial_{z}^{2}p-\partial_{z}(mp)\right],\quad p(\tau=0,z)=1,\\
\widetilde{q}(\tau)&=\frac{1}{2}\int_{-\infty}^{+\infty} dzp(\tau,z)\partial_{z}m(\tau,z),\label{eq:app:pankov:qt}\\
g^{\prime}(\tau)&=-\frac{1}{\tau}\frac{d}{d\tau}\phi^{\prime}(\widetilde{q}(\tau))\\
1&=\tau\phi^{\prime\prime}(\widetilde{q}(\tau))\int_{-\infty}^{+\infty} dz\,p(\tau,z)(\partial_{z}m(\tau,z))^{2}.\label{eq:app:pankov:marginalstability}
\end{align}

The following Ansatz then solves these equations in the parametric region $b \gg \tau \gg 1$:
\begin{equation}
    g^{\prime}(\tau)\simeq g_{0}/\tau^{3},\quad\widetilde{q}(\tau)\approx(g_{0}/\tau)^{n},\quad m(\tau,z)\approx m_{0}(z),\quad p(\tau,z)=(g_{0}/\tau)^{n}p_{0}(z),
\end{equation}
with universal functions $m_0(z)$ and $p_0(z)$ satisfying the following equations:
\begin{align}
zm_{0}^{\prime}+g_{0}\left(m_{0}^{\prime\prime}+2m_{0}m_{0}^{\prime}\right)&=0\\
zp_{0}^{\prime}-np_{0}-g_{0}\left(p_{0}^{\prime\prime}-2(m_{0}p_{0})^{\prime}\right)&=0.
\end{align}
with the boundary condition $m_0(z \to \pm \infty) = \mathrm{sign}(z)$.

At $z \to \infty$, the equation for $p_0(z)$ is then simplified and can be asymptotically solved:
\begin{equation}
    zp_{0}^{\prime}-np_{0}-g_{0}\left(p_{0}^{\prime\prime}-2 p_{0}^{\prime} \sign z\right)=0\Rightarrow p_{0}(z)\approx\frac{C}{(2g_{0})^{n/2}}H_{n}\left(\frac{|z|}{\sqrt{2g_{0}}}+\sqrt{2g_{0}}\right)\underset{z\gg1}{\approx}C\left(\frac{|z|}{g_{0}}\right)^{n},
\end{equation}
with the Hermite polynomial $H_n(z)$. Finally, the two constants that enter this scheme, $C$ and $g_0$, can be found from the equations which follow from Eqs. (\ref{eq:app:pankov:qt},\ref{eq:app:pankov:marginalstability}):
\begin{align}
\frac{g_{0}}{n}\int_{-\infty}^{+\infty} dzp_{0}(z)(m_{0}^{\prime}(z))^{2}&=1\\
\frac{1}{2}\int_{-\infty}^{+\infty} dzp_{0}(z)m_{0}^{\prime}(z)&=1
\end{align}

This scaling solution implies the following Ansatz for the form of the Coulomb gap in the distribution function of frozen fields, with $t \equiv T / E_C \equiv b^{-1}$
\begin{equation}
    p(1, y \ll 1) \approx C |y|^n + A t^n,
\end{equation}
which is Eq. \eqref{eq:CoulombGap} from the main text. This form implies vanishing of $\widetilde{q}(1)$ at zero temperature as:
\begin{equation}
\widetilde{q}(1)=\left(CI^{(1)}_{n}+A\right)t^{n},
\end{equation}
where 
\begin{equation}
I_{n}^{(1)}=\int_{0}^{\infty}\frac{z^{n}}{\cosh^{2}z}dz=2^{2(1-n)}(2^{n-1}-1)\Gamma(n+1)\zeta(n)=\begin{cases}
1, & n=0\\
\pi^{2}/12, & n=2\\
7\pi^{4}/240, & n=4
\end{cases}
\end{equation}
Furthermore, the value of constant $A$ can be related to $C$ utilizing the marginal stability criterion, which yields:
\begin{equation}
\frac{2}{n}\left(CI_{n}^{(1)}+A I_n^{(1)}\right)^{1/n-1}\left(I_{n}^{(2)}+A I_n^{(2)}\right)=1,
\end{equation}
where
\begin{equation}
I_{n}^{(2)}=\int_{0}^{\infty}\frac{z^{n}}{\cosh^{4}z}dz=\begin{cases}
2/3, & n=0\\
(\pi^2-6)/18, & n=2
\end{cases}
\end{equation}
For the low-temperature behavior of the entropy it implies:
\begin{equation}
S/\nu_{0}E_{C}=2\left[CI_{n}^{(3)}+AI_{0}^{(3)}-\frac{n}{n+1}\left(CI_{n}^{(1)}+A\right)^{1+1/n}\right]t^{n+1},
\end{equation}
with
\begin{equation}
I_{n}^{(3)}=\int_{0}^{\infty}z^{n}(\ln2\cosh z-z\tanh z)dz=\frac{I_{n+2}^{(1)}}{n+1}
\end{equation}

\begin{figure}[ht]
    \centering
    \includegraphics[width=\textwidth]{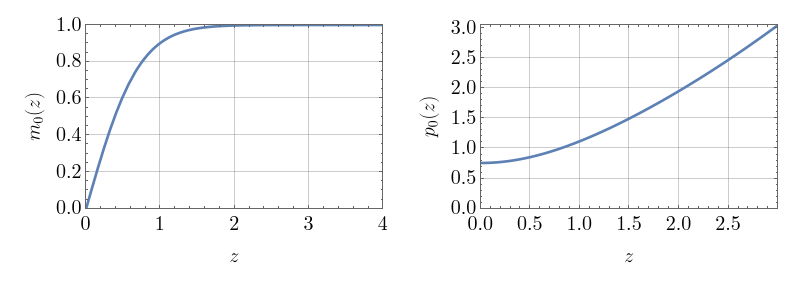}
    \caption{Pankov scaling functions $m_0(z)$ and $p_0(z)$ for 3D Coulomb glass $n=2$}
    \label{fig:app:Pankov}
\end{figure}

We have implemented the numerical self-consistent solution of scaling equations for $n=2$, corresponding to the three-dimensional Coulomb glass, and have obtained the following values:
\begin{equation}
C \approx 0.32673,\quad g_0 \approx 1.12303,
\end{equation}
and the scaling functions $m_0(z)$ and $p_0(z)$ are plotted on Fig. \ref{fig:app:Pankov}. This implies:
\begin{align}
A&\approx2.29761\\
\widetilde{q}(1)&\approx 2.56633\,(T/E_C)^2\\
p(1,y)&\approx 0.32673\,y^2+2.29761\,(T/E_{C})^{2},\\
S/P_{0}E_{C}&\approx1.65746\,(T/E_C)^3,
\end{align}
which gives Eqs. (\ref{eq:CoulombGap},\ref{eq:G:Ent:LowT}) from the main text.

\section{Properties of superconducting phase with Coulomb corrections}
\subsection{Free energy}
\label{sec:appendix:SC}

This Appendix is devoted to detailed analytical analysis of the free energy of the superconducting phase with Coulomb corrections.

The first contribution is ``local'', given by Eq. \eqref{eq:SC:Floc}. We subtract the ``band'' contribution, which corresponds to the same integral at $T = 0$ and $\Delta = 0$:
\begin{equation}
{\cal F}_{b,0}=-n_{0}\int_{-\infty}^{+\infty} d\xi\,P_{0}(\xi)|\xi|=-\sqrt{\frac{2}{\pi}}n_{0}W,
\end{equation}
and obtain an integral with the main contribution coming from the vicinity of the Fermi energy, allowing us to replace the distribution function by a constant:
\begin{multline}
    {\cal F}_{\text{loc}}-{\cal F}_{b,0}\simeq
    \frac{P_{0}|\Delta|^{2}}{\lambda}-\nu_{0}\int_{-\omega_{D}}^{\omega_{D}}d\xi\left(\sqrt{\xi^{2}+|\Delta|^{2}}-|\xi|\right)-\nu_{0}T\int_{-\infty}^{\infty}d\xi\ln\left(1+e^{-2\beta\sqrt{\xi^{2}+|\Delta|^{2}}}\right)\\
= \nu_{0}\Delta^{2}\left(\frac{1}{2}\ln\frac{|\Delta|^{2}}{e\Delta_{0}^{2}}-\eta(\beta|\Delta|)\right),
\end{multline}
with the dimensionless function $\eta(z)$ defined in Eq. \eqref{eq:eta}, which has following asymptotic behavior:
\begin{equation}
\eta(z)\approx\begin{cases}
\sqrt{\pi}e^{-2z}/z^{3/2}, & z\gg1\\
\pi^{2}/12z^{2}-\ln(\pi e^{-\gamma}/2z)-1/2, & z\ll1
\end{cases}
\end{equation}

The quenched fluctuations of ``magnetization'' are given by Eq. \eqref{eq:SC:Q0}. Adding and subtracting unity, we can rewrite it as:
\begin{equation}
    Q_{0}=1 - \int_{-\infty}^{+\infty} d\xi\,P_{0}(\xi)(1 - m^{2}(\xi))
\end{equation}
The remaining integral is again dominated by the vicinity of the Fermi-surface. This brings us to the following form:
\begin{equation}
    Q_{0} \approx 1-P_{0}|\Delta|\,q_{0}(\beta|\Delta|),
\end{equation}
with the dimensionless function 
\begin{equation}
q_{0}(z)=\int_{-\infty}^{\infty}dy\left[1-\frac{y^{2}}{y^{2}+1}\tanh^{2}\left(z\sqrt{y^{2}+1}\right)\right]\approx\begin{cases}
\pi+2\sqrt{\pi}e^{-2z}/z^{3/2} & z\gg1\\
2/z, & z\ll1
\end{cases}
\end{equation}

The average dynamic susceptibility is given by Eq. \eqref{eq:SC:Q1}. The integral is already dominated by the vicinity of the Fermi surface. At bosonic Matsubara frequency $\omega = 2 \pi n T$, we obtain:
\begin{equation}
\label{eq:app:SC:Q1}
\widetilde{Q}_{1}(\omega = 2 \pi n T)\approx2 P_0\widetilde{q}_{n}(\beta|\Delta|),
\end{equation}
with the set of dimensionless functions
\begin{equation}
\widetilde{q}_{n\neq0}(z)=\frac{1}{2}\int_{-\infty}^{\infty}dy\frac{\tanh z\sqrt{y^{2}+1}}{\sqrt{y^{2}+1}\left(y^{2}+1+\left(\pi n/z\right)^{2}\right)}\approx\begin{cases}
\theta_{n}(z)/\sinh\theta_{n}(z), & z\gg1\\
c_{n}z^{2}, & z\ll1
\end{cases},
\end{equation}
and $\widetilde{q}_0(z) \equiv 1$. Here we have denoted
\begin{equation}
\sinh\frac{\theta_n(z)}{2}=\frac{\pi n}{z}\quad c_{n}=\int_{0}^{\infty}dx\frac{\tanh x}{x\left(x^{2}+\pi^{2}n^{2}\right)}
\end{equation}

Finally, the Coulomb contribution \eqref{eq:SC:FC} can be brought to the dimensionless form as follows:
\begin{equation}
{\cal F}_{C}=-n_{0}E_{C}+\nu_{0}E_{C}|\Delta|\left(q_{0}(\beta|\Delta|)-\frac{4}{3}c(\beta|\Delta|)\right),
\end{equation}
where we have introduced yet another dimensionless function
\begin{equation}
c(z)=\frac{1}{z}\sum_{n\in\mathbb{Z}}\widetilde{q}_{n}^{3/2}(z)\approx\begin{cases}
C_{1}, & z\gg1,\\
1/z+C_{2}z^{2}, & z\ll1,
\end{cases},
\end{equation}
with constants
\begin{equation}
C_{1}\approx\int_{-\infty}^{\infty}\frac{d\theta}{2\pi}\cosh\frac{\theta}{2}\left(\frac{\theta}{\sinh\theta}\right)^{3/2}\approx0.896167,\quad C_{2}=2\sum_{n=1}^{\infty}c_{n}^{3/2}\approx0.249127,
\end{equation}

Notably, in the limit $z \to \infty$, both functions $q_0(z)$ and $c(z)$ saturate at a constant. As the result, in the $T \to 0$ limit, the Coulomb contribution is non-analytic in $\Delta$:
\begin{equation}
{\cal F}_{C}(T\to0)=-n_{0}E_{C}+C\,\nu_{0}E_{C}|\Delta|,
\end{equation}
with 
\begin{equation}
C = \pi - \frac{4}{3} C_1 \approx 1.946703
\end{equation}

\subsection{Stability}
\label{sec:appendix:SCStability}

Finally, we switch to the analysis of the possibility of the soft instability of the superconducting state towards appearance of the glass order parameter on top of it --- as related to the question of possible \emph{coexistence} of superconducting and glass order parameters in the present problem. Such instability appears if the ``glass susceptibility'' $\chi_{\text{SG}}$, defined as:
\begin{equation}
\chi_{\text{SG}}=n_{0}^{-1}\Phi^{\prime\prime}(\beta\widetilde{Q})\int_{-\infty}^{+\infty}d\xi P_{0}(\xi)\left(\frac{\partial m}{\partial\xi}\right)^{2},
\end{equation}
becomes larger than unity (i.e. at the highest temperature where the marginal stability criterion \eqref{eq:app:MarginalStability} is fulfilled). Here $\widetilde{Q} = \widetilde{Q}_1(\omega_n = 0) = 2P_0$ is given by Eq.~\eqref{eq:app:SC:Q1}, and $m(\xi)$ is given by Eq.~\eqref{m1}. In the dimensionless units $z \equiv \beta \Delta$ and $y \equiv \xi / \Delta$, it reads:
\begin{equation}
\label{eq:app:ChiSG}
\chi_{\text{SG}}\left(T,E_{C},\Delta\right)=\frac{E_{C}}{\Delta}\int_{0}^{\infty}dy\,\left(\frac{\partial}{\partial y}\left[\frac{y}{\sqrt{y^{2}+1}}\tanh\left(z\sqrt{y^{2}+1}\right)\right]\right)^{2}
\end{equation}

\begin{figure}[ht]
    \centering
    \includegraphics[width=0.5\textwidth]{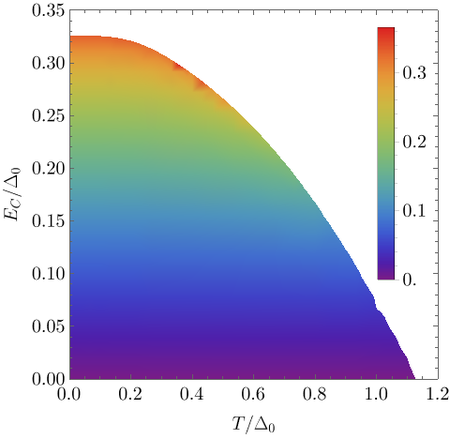}
    \caption{Spin glass susceptibility $\chi_{\text{SG}}$, calculated on top of the superconducting state.}
    \label{fig:GlassSusc}
\end{figure}

We have calculated this quantity numerically everywhere inside the superconducting region on the phase diagram, and the result is presented on Fig.~\ref{fig:GlassSusc}. One can see that everywhere in the superconducting phase, the susceptibility is well below unity, i.e. the superconducting state is everywhere stable towards appearance of the glass order parameter.

\end{appendix}

\bibliography{refs}

\end{document}